\documentclass[twocolumn,aps,preprint,10pt,superscriptaddress,floatfix,showpacs]{revtex4-1}
\usepackage{graphics}
\usepackage{dcolumn}
\usepackage{bm}
\usepackage{epsfig}
\usepackage{color}
\usepackage{epic}
\usepackage{hyperref}

\begin{document}

\title{Pulsed laser deposition growth of heteroepitaxial YBa$_2$Cu$_3$O$_7$/La$_{0.67}$Ca$_{0.33}$MnO$_3$ superlattices on NdGaO$_3$ and Sr$_{0.7}$La$_{0.3}$Al$_{0.65}$Ta$_{0.35}$O$_3$  substrates}



\author{ V. K. Malik}
\altaffiliation[Now at]{ Department of Chemical Engineering and Material Science, Univerisity of California, Davis, 95616 CA, USA}
\affiliation{University of Fribourg, Department of Physics and Fribourg Centre for Nanomaterials, 
Chemin du Mus\'{e}e 3, CH-1700 Fribourg, Switzerland
}
\author{I. Marozau}
\affiliation{University of Fribourg, Department of Physics and Fribourg Centre for Nanomaterials, 
Chemin du Mus\'{e}e 3, CH-1700 Fribourg, Switzerland
}
\author{S. Das}
\affiliation{University of Fribourg, Department of Physics and Fribourg Centre for Nanomaterials, 
Chemin du Mus\'{e}e 3, CH-1700 Fribourg, Switzerland
}
\author{B. Doggett}
\affiliation{University of Fribourg, Department of Physics and Fribourg Centre for Nanomaterials, 
Chemin du Mus\'{e}e 3, CH-1700 Fribourg, Switzerland
}
\author{D. K. Satapathy}
\affiliation{University of Fribourg, Department of Physics and Fribourg Centre for Nanomaterials, 
Chemin du Mus\'{e}e 3, CH-1700 Fribourg, Switzerland
}
\author{M. A. Uribe-Laverde}
\affiliation{University of Fribourg, Department of Physics and Fribourg Centre for Nanomaterials, 
Chemin du Mus\'{e}e 3, CH-1700 Fribourg, Switzerland
}
\author{N. Biskup}
\affiliation{Universidad Complutense de Madrid, Madrid 28040, Spain}
\author{M. Varela}
\affiliation{Universidad Complutense de Madrid, Madrid 28040, Spain}
\affiliation{Oak Ridge National Laboratory, Oak Ridge, TN 37831 USA}
\author{C. W. Schneider}
\affiliation{Materials Group, Paul Scherrer Institut, CH-5232 Villigen, Switzerland}
\author{C. Marcelot}
\affiliation{Laboratory for Neutron Scattering Paul Scherrer Institut, CH-5232 Villigen, Switzerland}
\author{J. Stahn}
\affiliation{Laboratory for Neutron Scattering Paul Scherrer Institut, CH-5232 Villigen, Switzerland}
\author{C. Bernhard}
\email{christian.bernhard@unifr.ch}
\affiliation{University of Fribourg, Department of Physics and Fribourg Centre for Nanomaterials, 
Chemin du Mus\'{e}e 3, CH-1700 Fribourg, Switzerland
}


\date{\today}

\begin{abstract}
Heteroepitaxial superlattices of [YBa$_2$Cu$_3$O$_7$(n)/La$_{0.67}$Ca$_{0.33}$MnO$_3$(m)]$_\mathrm{x}$ (YBCO/LCMO), where n and m are the number of YBCO and LCMO monolayers and x the number of bilayer repetitions, have been grown with pulsed laser deposition on NdGaO$_3$ (110) and Sr$_{0.7}$La$_{0.3}$Al$_{0.65}$Ta$_{0.35}$O$_{3}$ (LSAT) (001). These substrates are well lattice matched with YBCO and LCMO and, unlike the commonly used SrTiO$_3$, they do not give rise to complex and uncontrolled strain effects at low temperature. The growth dynamics and the structure have been studied \emph{in-situ} with reflection high energy electron diffraction (RHEED) and \emph{ex-situ} with scanning transmission electron microscopy (STEM), x-ray diffraction and neutron reflectometry. The individual layers are found to be flat and continuous over long lateral distances with sharp and coherent interfaces and with a well- defined thickness of the individual layer. The only visible defects are antiphase boundaries in the YBCO layers that originate from perovskite unit cell height steps at the interfaces with the LCMO layers. We also find that the first YBCO monolayer at the interface with LCMO has an unusual growth dynamics and is lacking the CuO chain layer while the subsequent YBCO layers have the regular Y-123 structure. Accordingly, the CuO$_2$ bilayers at both the LCMO/YBCO and the YBCO/LCMO interfaces are lacking one of their neighboring CuO chain layers and thus half of their hole doping reservoir. Nevertheless, from electric transport measurements on a superlattice with n=2 we obtain evidence that the interfacial CuO$_2$ bilayers remain conducting and even exhibit the onset of a superconducting transition at very low temperature. Finally, we show from dc magnetization and neutron reflectometry measurements that the LCMO layers are strongly ferromagnetic.
\end{abstract}


\maketitle

\section{Introduction}
\label{intro}
The concept of combining materials with competing orders in the form of artificially grown multilayers and superlattices (SLs) is a very popular approach to obtain novel materials with modified or even with entirely new physical properties. These properties can be readily tuned with external parameters like strain, electric or magnetic fields\cite{Jiang:1995p19104,Mercaldo96,Obi:1999p19214,Pena:2005p13371,Pena:2006p3494,Stahn:2005p13350,CHAKHALIAN:2006p3735,Hoppler:2009p18981}. Complex transition metal oxides provide an ideal test bed, since they offer a rich spectrum of individual physical properties\cite{dagotto59,Fath99,Hanguri04}. Also, thanks to their similar lattice structure and chemical compatibility many of them can be readily combined to grow high quality thin film heterostructures. Recently, there has been a tremendous progress in the growth of oxide thin films with the technique of pulsed laser deposition (PLD), which has made it possible to prepare oxide multilayers and superlattices with very high structural quality and with interfaces that are flat and chemically sharp on the atomic scale\cite{Ohtomo02,Seo10,habermeier:2001p13348,Christen:2008p3677,Mandal08,Martin:2007p19303,Garcia:2010p19317,Singh:2006p19323}. A prominent example are the LaAlO$_3$/SrTiO$_3$ heterostructures which host a two dimensional interfacial electron gas whose mobility can be very high and which can even become superconducting\cite{Ohtomo:2004p19345,Thiel:2006p19346,Reyren:2007p19347}.\\
These oxide multilayers also provide unique possibilities to combine the mutually exclusive ferromagnetic and superconducting orders. As compared to their conventional counterparts which have been already extensively studied\cite{Bulaevskii:1985p19348,Buzdin:2005p318}, the oxide-based superconductor/ferromagnet (SC/FM) multilayers offer some appealing properties, like the very high superconducting transition temperature, $T_\mathrm{C}$, of the cuprates or the fully spin polarized state of charge carriers the manganites. The latter are well known for their so-called colossal magneto-resistance (CMR) effect\cite{Jin:1994p18761}. However, while the physical properties of the conventional SC/FM multilayers are fairly well understood\cite{Eschrig:2011p19001}, there is presently no consensus on the corresponding properties of the oxide-based counterparts.\\
The most commonly investigated systems are heterostructures where the cuprate high temperature superconductor YBa$_2$Cu$_3$O$_7$ (YBCO) is combined with the ferromagnetic manganites La$_{2/3}$Ca$_{1/3}$MnO$_3$ (LCMO) or La$_{2/3}$Sr$_{1/3}$MnO$_3$ (LSMO)\cite{Pena:2005p13371,Pena:2006p3494,habermeier:2001p13348,Mandal08}. Previous studies revealed a number of interesting phenomena, like an anomalous suppression of the free carrier response\cite{Holden:2004p13349}, a giant magnetoresitance effect\cite{Pena:2005p13371}, an unusually large photo-doping effect on $T_\mathrm{C}$\cite{Pena:2006p3494}, an antiphase magnetic proximity effect\cite{Bergeret:2005p11119} with an induced ferromagnetic moment of Cu\cite{Stahn:2005p13350,CHAKHALIAN:2006p3735}, and a giant modulation of the ferromagnetic order in the LCMO layers that is induced by the superconducting transition of the YBCO layers\cite{Hoppler:2009p18981}. While the wealth of these unusual phenomena makes these YBCO/LCMO heterostructures appealing candidates for applications in future electronic devices, the understanding of the relevant physical parameters and interactions in these materials requires further intense research efforts. This circumstance is related to the complex structural properties of these oxides and to the extreme sensitivity of their physical properties even to minute structural or compositional changes.\\
A remarkable example was reported in Ref. \cite{Hoppler:2009p18981} where the superconductivity-induced modulation of the ferromagnetic order was shown to depend on the strain condition ensuing from the way the SrTiO$_3$ substrate was mounted on the sample holder. These SrTiO$_3$ substrates were in fact found to develop a pronounced buckling of the near surface region in the context of some structural phase transitions below 105\,K. The strain pattern that arises from this buckling was transmitted into the YBCO/LCMO superlattice deposited on top\cite{Hoppler:2008p9625}. By applying a weak external pressure during cooling, this buckling pattern could be strongly modified and dramatic changes of the magnetic properties of the YBCO/LCMO superlattices could be obtained\cite{Hoppler:2009p18981}. This effect is another clear manifestation that the magnetic properties of the manganites are extremely versatile and can be largely modified even by small perturbations.  But it also provides a clear warning that the structural properties of these YBCO/LCMO heterostructures and of the substrates on which they are grown must be better understood and controlled. This is one of the motivations for the present work, which is concerned with the growth of YBCO/LCMO superlattices on NdGaO$_3$ and LSAT substrates which do not undergo any phase transitions below 300\,K which could give rise to complex structural changes.\\
Another important aspect of these superlattices that directly relates to their physical properties (in particular to the coupling between the superconducting and ferromagnetic order parameters) concerns the termination of the YBCO monolayers at the YBCO/LCMO and the LCMO/YBCO interfaces. In previous works different results have been reported for YBCO/LCMO heterostructures on SrTiO$_3$ substrates. For superlattices grown with high oxygen pressure sputtering Varela and coworkers reported a layer sequence of YBCO-CuO$_2$-Y-CuO$_2$-BaO-MnO$_2$ ÐLCMO for the YBCO/LCMO interface which implies that the YBCO monolayers next to the interfaces are missing the CuO chain layer\cite{Varela:2003p17134}. This finding should have important consequences for the electronic state of the neighboring CuO$_2$ bilayers for which the CuO chains act as a charge reservoir. Accordingly, these interfacial CuO$_2$ bilayers may be strongly underdoped and possibly they may be even insulating, thus acting as barriers that strongly reduce the proximity coupling between the superconducting and ferromagnetic orders. A notably different layer stacking sequence has been reported for a PLD grown YBCO/LCMO/YBCO trilayer on SrTiO$_3$ by Zhou and coworkers\cite{Zhang:2009p14937}. They observed two different interface configurations where the CuO chains are present at the LCMO/YBCO interface with the layer sequence LCMO-(La,Ca)O-MnO$_2$-BaO-CuO-BaO-CuO$_2$-Y-CuO$_2$-YBCO,  whereas the CuO chains and even half of the CuO$_2$ bilayer are missing at the YBCO/LCMO interface with the layer sequence YBCO-BaO-CuO$_2$-(La,Ca)O-MnO$_2$-LCMO. The second goal of our present work therefore has been to investigate the interfacial termination in YBCO/LCMO heterostructures where the growth and the thickness of the individual layers is controlled on the monolayer scale with \emph{in-situ} reflection high energy electron diffraction (RHEED).
\section{Experiment}
\label{expt}
The PLD deposition system (SURFACE-TEC Gmbh) consists of an ultra-high vacuum chamber (10$^{-9}$ mbar base pressure), a KrF excimer laser for the ablation (Coherent GmbH, COMPexPro 205\,F) with a wavelength of 248\,nm and a pulse duration of 25\,ns, an infrared laser (JENOPTIK, JOLD-140-CAXF-6A) with a wavelength of 807\,nm for the substrate heating and an infrared pyrometer for the control of the substrate temperature. The infrared laser radiation is focused on the backside of the substrate holder to which the substrate is glued with silver paint. A special design of the substrate holder, which has only a weak thermal coupling to the outer parts of the sample holder system, ensures that the substrate is uniformly heated while the remaining part of the system, which also becomes coated during the PLD growth, remains well below 100$^{\circ}$\,C. This helps to avoid cross contamination by thermal reevaporation which is a common problem with resistive heaters. The exchange of the ablation targets is enabled with a computer controlled rotation system.\\
The system also contains an \emph{in-situ} reflection high energy electron diffraction (RHEED) system with a 30\,keV electron gun (RDA-002\,G, R-DEC co. Ltd.), a two-stage differential pumping unit and a stainless steel tube with a small aperture of about 0.5\,mm diameter close to the substrate which ensure that the RHEED measurements can be performed at high background gas pressures in the growth chamber up to 50\,mTorr. The RHEED diffraction pattern is imaged with a phosphorous screen which is monitored with a CCD camera whose output is analyzed with a commercial software package (K-Space).\\
The [YBa$_2$Cu$_3$O$_7$(n)/La$_{2/3}$Ca$_{1/3}$MnO$_3$(m)]$_\mathrm{x}$ superlattices with n unit cells of YBCO per layer and m unit cells of LCMO and x repetitions of the YBCO(n)/LCMO(m) bilayers were grown with pulsed laser deposition (PLD) on NdGaO$_3$ (NGO) (110) and Sr$_{0.7}$La$_{0.3}$Al$_{0.65}$Ta$_{0.35}$O$_3$ (LSAT) (001) substrates as specified in Table \ref{summarySLs_table}. 
\begin{table*}
\hfill
  \begin{center}
    \begin{tabular}{|m{2cm}|m{2cm}|m{2cm}|m{2cm}|m{2cm}|m{2cm}|m{2cm}|m{2cm}|}
      \hline
{$Sample\:no.$}	 &{$SL-287$}	& {$SL-288$} & {$SL-327$} & {$SL-428$} &{$SL-427$} &{$SL-448$} & {$SL-447$} \cr
      \hline
{Substrate}			&{NGO}			&{LSAT}  &{NGO} &{LSAT}  &{LSAT}  &{LSAT}  &{LSAT} \cr
      \hline
{n}		& 8			& 8 & 5 & 1 & 2 & 3 & 4  \cr
      \hline
      {m}		& 28			& 28 & 13 & 12 & 12 & 12 & 12  \cr
      \hline     
       \footnotesize{$T_\mathrm{C}$($\textrm{K}$)}		& 78			& 75 & 60 & - & 7 & 45 & 60  \cr
      \hline 
      \footnotesize{$T_\mathrm{Curie}$($\textrm{K}$)}		& 225			& 225 & 200 & 200 & 200 & 200 & 200 \cr
      \hline            
    \end{tabular}
    \caption{List of the number of monolayers and the superconducting and ferromagnetic transition temperatures of the investigated [YBa$_2$Cu$_3$O$_7$(n)/La$_{2/3}$Ca$_{1/3}$MnO$_3$(m)]$_\textrm{x}$ superlattices.}
    \label{summarySLs_table}
  \end{center}
\end{table*}
The growth parameters have been obtained by optimizing the growth conditions for the individual YBCO and LCMO films with a thickness of about 10\,nm. The growth of the superlattices was always started with a YBCO layer and finished with a LCMO layer. The substrates of lateral size 10$\times$10\,mm$^{2}$ were heated to 825\,$^{\circ}$C in an oxygen partial pressure of 0.34\,mbar at which they were annealed for 30 minutes for degassing and eventually a curing of the mechanically polished surface. Subsequently, the YBCO/LCMO superlattices were grown at the same temperature and oxygen partial pressure by laser ablation from very dense and stoichiometric sintered, ceramic targets (pi-KEM, 99.9\% purity). These targets were placed approximately 6\,cm below the substrate and were rotated and toggled translationally to achieve a homogeneous ablation of the entire target surface (19.6\,cm$^2$). Using a mechanically operated shutter in front of the substrate, the target surface was conditioned (preablated) prior to the deposition of each individual layer as to avoid cross contamination. The laser fluence and frequency were 1.5 to 2.0\,J/cm$^2$ and 7\,Hz, respectively. The laser beam was focused on the target with a system of two optical lenses, two mirrors and one mechanically adjustable slits to achieve a well-defined and homogenous spot size of about 6\,mm$^{2}$. After the deposition, the samples were cooled to 700\,$^{\circ}$C at a rate of about 10\,$^{\circ}$C/min while the oxygen partial pressure was increased to 1\,bar. Subsequently, the temperature was decreased to 485\,$^{\circ}$C at a rate of about 30\,$^{\circ}$C/min and the samples were annealed for one hour (\emph{in-situ} post-deposition annealing). Finally the samples were slowly cooled down to room temperature and extracted from the PLD chamber. The \emph{ex-situ} post-deposition annealing was performed in a separate furnace in a gas flow of pure oxygen (100\,ml/min) for 12 hours at 485$^{\circ}$\,C with subsequent slow cooling to room temperature.\\
The NdGaO$_3$(NGO) substrates crystallize in the GdFeO$_3$ type orthorhombic structure with lattice parameters a=5.428\,\AA, b=5.498\,\AA, and c=7.708\,\AA. Substrates with (110) oriented surfaces were used whose in-plane \emph{pseudo-cubic} lattice parameters amount to $a_\mathrm{p}$=3.863\,\AA \,and $b_\mathrm{p}$=3.854\,\AA, respectively. The resulting lattice mismatch with respect to YBCO and LCMO is very small ($<$1\%), it gives rise to a weakly compressive strain condition. With x-ray diffraction measurements we confirmed that the NGO substrates are of high structural quality and do not undergo any structural phase transition below room temperature. 
The Sr$_{0.7}$La$_{0.3}$Al$_{0.65}$Ta$_{0.35}$O$_3$ (LSAT) substrates have cubic crystal symmetry at room temperature with a lattice constant of 3.868\,\AA. They have a very small lattice mismatch with respect to YBCO and LCMO ($<$0.5\%). LSAT is known to undergo a structural transition around 150\,K from cubic to tetragonal symmetry\cite{Chakoumakos:1998p18967}. However, this transition involves only a very small lattice distortion and it does not lead to significant strain effects on the thin films on top as is the case for the SrTiO$_3$ substrates.\\
During the PLD growth the \emph{in-situ} high pressure RHEED setup was used to study the growth dynamics of different layers of the YBCO/LCMO superlattices. A clear modulation of the intensity of the RHEED peaks was observed for the YBCO and the LCMO layers. These growth oscillations were used to monitor the layer by layer growth of the individual YBCO and LCMO layers. Information about the in-plane periodicity and the roughness of the surface layer has been obtained from the off specular diffraction signal.\\
The structural analysis of the SLs and their substrates was performed with x-ray diffraction (XRD), with specular neutron reflectivity, as well as with scanning transmission electron microscopy (STEM). For the XRD measurements we used a Seifert diffractometer with a 0.5\,mm receiving slit. This system was equipped with a x-ray mirror, a four crystal Ge (220) monochromator and it utilizes Cu-K$_\mathrm{\alpha}$ radiation with a wavelength of 0.154\,nm. All XRD scans were performed at room temperature. The polarized neutron reflectivity measurements were performed at the Morpheus instrument at the quasi-continuous neutron spallation source (SINQ) of the Paul-Scherer Institut (PSI) in Villigen, Switzerland. The neutron wavelength of was chosen with the help of a monochromator which yields a small wavelength dispersion of $\Delta \lambda/\lambda \sim 1\%$.  To obtain a high enough neutron flux at the sample position, the slit before the sample was kept at 0.6 mm. The spin-polarization of the neutron beam was obtained with a multi-layer polarizer mirror and a Mesay-type spin flipper. No analyzer mirror was used for the reflectometry measurements reported here. Helmholtz coils were used to create an external magnetic field of up to 1000\,Oe that was oriented perpendicular to the scattering plane and parallel to the film surface. Further details on the experimental setup are given in Ref.\cite{Hoppler09}.  The reflection curves have been simulated using the program superfit \cite{Ruhm:1999p17046}.\\
The high resolution scanning transmission electron microscopy measurements where performed at the aberration corrected Nion ultraSTEM 100 at Oak Ridge National Laboratory, operated at 100\,kV and equipped with a fifth order aberration corrector and a Gatan Enfina electron energy loss spectrometer (EELS). Cross sectional experiments were prepared by conventional grinding and Ar ion milling.\\
The electric resistivity and dc magnetization measurements have been carried out using the four-point probe resistivity and the vibrating sample magnetometer (Model P525) options of a physical properties measurement system (PPMS) of Quantum Design (Model QD6000). For the magnetization measurements a small piece with a rectangular shape was cut from the corner of the sample.\\
\section{Results and Discussion}
\label{res_diss}
\subsection{Structural Characterization}
\label{stuct_char}
\subsubsection{In-situ \emph{RHEED control}}
\label{RHEED}
The technique of \emph{in-situ} high pressure RHEED is widely used for the online monitoring of the surface structure and the growth dynamics of oxide thin films. It has already been extensively applied to investigate the growth of YBCO and manganite thin films\cite{Rijnders:1997p16077,Huijben:2008p17059}, however, to the best of our knowledge it has not yet been used to study the growth of YBCO/LCMO superlattices. In the following we show that these \emph{in-situ} RHEED measurements provide new insight into the growth mechanism of these superlattices. In particular, they reveal that the growth of the first YBCO monolayer on LCMO proceeds differently from the one of the following YBCO monolayers.\\
      \begin{figure} [h!]
             \begin{picture}(100,100)
          \put(-70,-5){\rotatebox{0}{\includegraphics[height=112\unitlength]{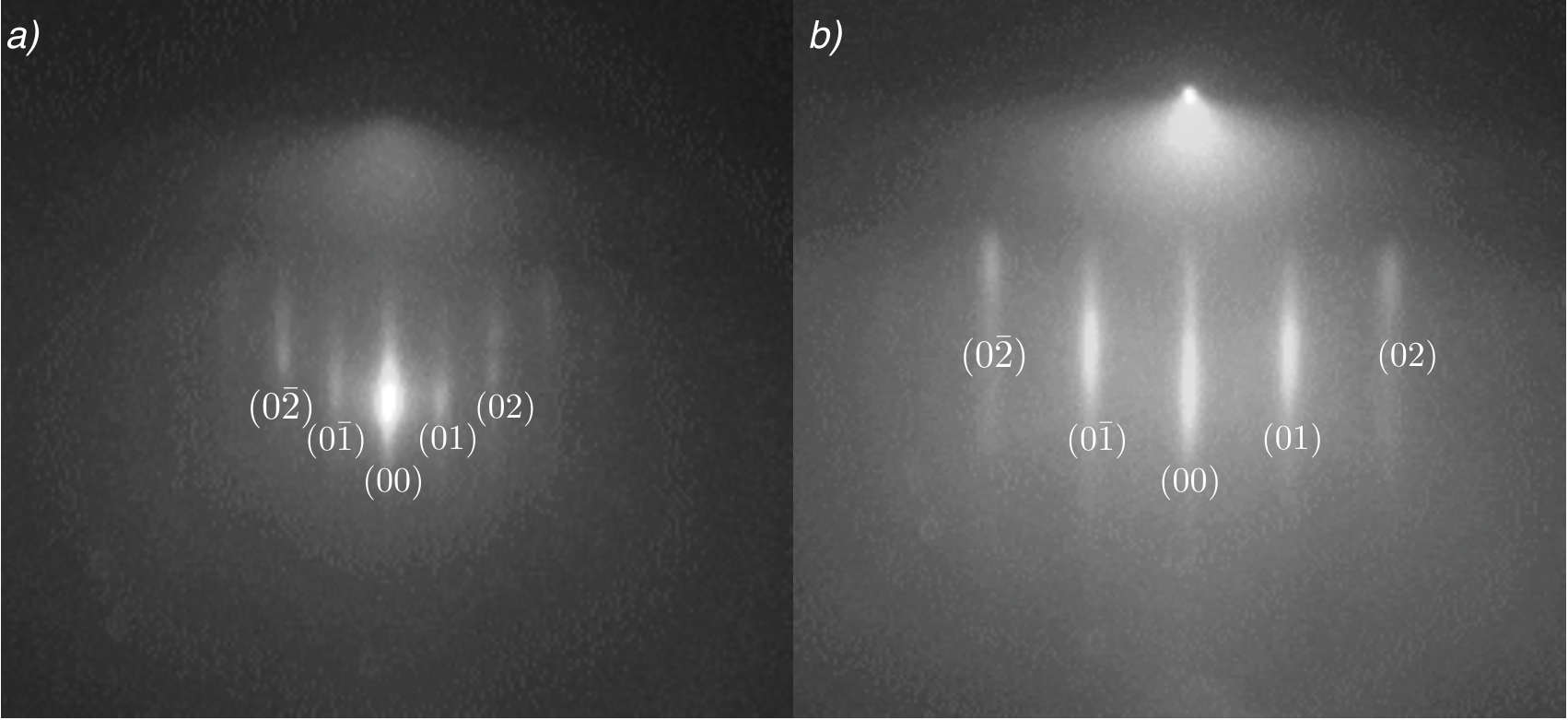}}}
        \end{picture}
        \vspace{0ex}
        \caption{ \label{rheed_pattern}
      \emph{In-situ} RHEED pattern as obtained from (a) the surface of the bare NGO substrate after heating for 30 minutes at 825\,$^{\circ}$C, and (b) the surface of the final LCMO layer on top of the [YBCO(n=8)/LCMO(m=27)]$_{10}$ superlattice (SL-287) with a total thickness of about 200\,nm}
      \end{figure}
Fig.~\ref{rheed_pattern}~$(a)$ displays a representative RHEED pattern of the bare NGO substrate after it has been annealed for 30 minutes at 825\,$^{\circ}$C.  The pattern reveals clear diffraction spots up to the 2$^\mathrm{nd}$ order (02) that are characteristic of a well-ordered, flat surface. Fig.~\ref{rheed_pattern}~$(b)$ shows the corresponding RHEED pattern from the surface of the final LCMO layer on the top of the [YBCO(n=8)/LCMO(m=28)]$_{10}$ superlattice (SL-287) with a total thickness of about 200\,nm. It reveals a clear pattern of elongated diffraction peaks. These so-called streaks arise from finite size effects and are characteristic of a two-dimensional surface structure with a limited lateral correlation length. These RHEED data confirm that the quality of the surface remains fairly high, even after the growth of ten YBCO/LCMO bilayers.\\
      \begin{figure} [h!]
       \begin{picture}(500,500)
          \put(0,0){\rotatebox{0}{\includegraphics[height=500\unitlength]{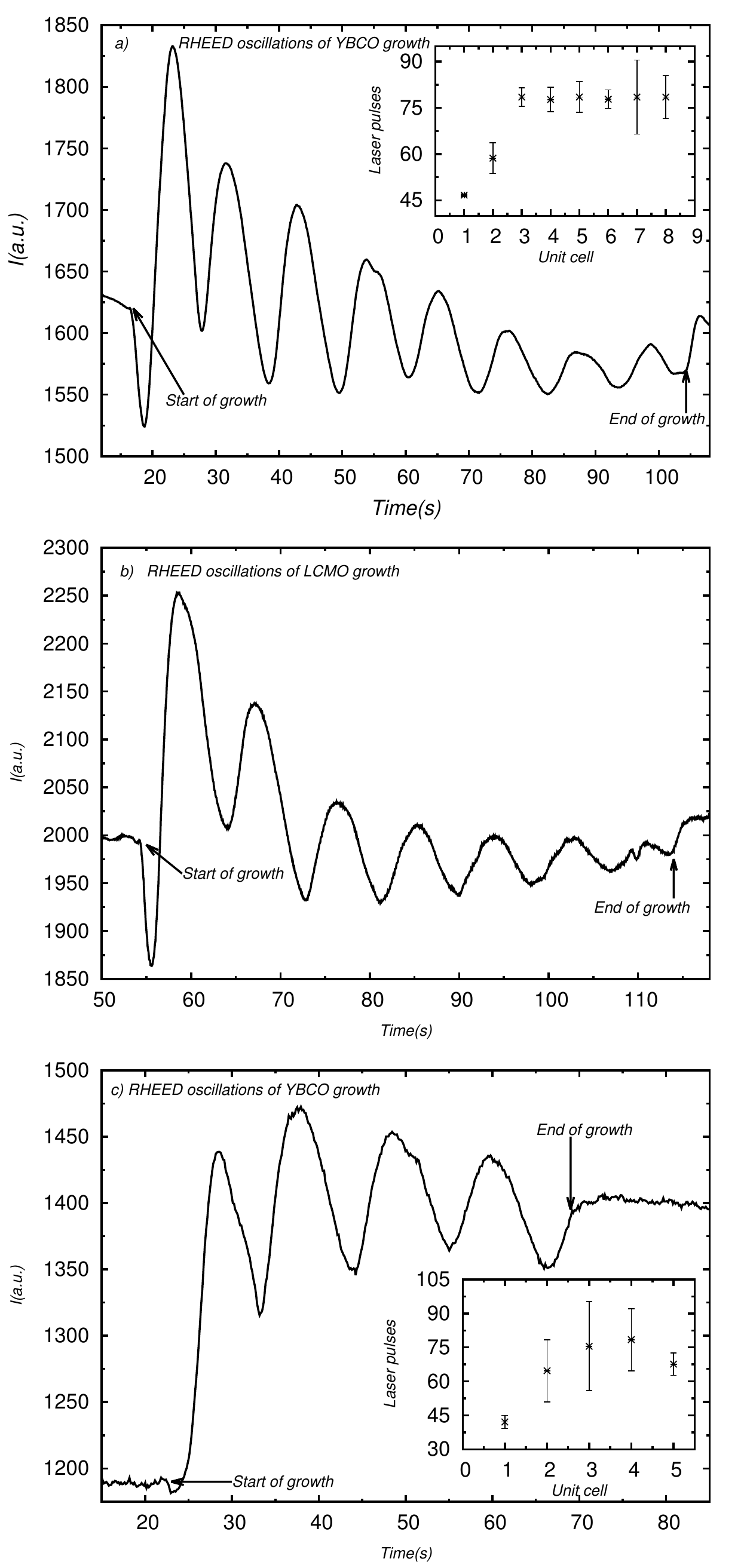}}}
        \end{picture}
        \vspace{0ex}
        \caption{ \label{rheed_oscillations}
      Time evolution of the intensity of the (00) peak in the specular RHEED pattern during the growth of (a) 8 monolayers of YBCO on LCMO, (b) 7 monolayers of LCMO on YBCO, and (c) 5 monolayers of YBCO on LCMO}
      \end{figure}
We also studied the growth dynamics of the individual YBCO and LCMO layers during the deposition of the superlattices by monitoring the time evolution of the (00) peak in the specular RHEED pattern. The peak intensity was found to exhibit a well resolved oscillatory time dependence. The intensity maxima occur at the time when the roughness of the surface layer is minimal. They can be used to keep track of the number of completed YBCO or LCMO monolayers that have been grown. Figs.~\ref{rheed_oscillations}~$(a)$ and $(b)$ show representative examples for the time evolution of the RHEED signal during the growth of a YBCO layer on top of LCMO and of a LCMO layer on YBCO, respectively. In both cases the growth oscillations are only moderately damped and they can be used to monitor the number of the YBCO and LCMO monolayers that have been grown. After the end of the deposition of each YBCO or LCMO layer the RHEED signal exhibits a clear recovery. The characteristic RHEED pattern, the pronounced growth oscillations, and the intensity recovery at the end of the growth are all indicative of a two-dimensional growth mode and of the formation of coherent and fairly flat interfaces between the YBCO and LCMO layers.\\
The RHEED data in Fig.~\ref{rheed_oscillations}~$(a)$ furthermore provide evidence that for the first YBCO monolayer on top of LCMO (or even on the NGO substrate) the growth dynamics is rather different from the one of the subsequent YBCO monolayers. This is evident in the inset of Fig.~\ref{rheed_oscillations}~$(a)$ which shows the number of laser pulses that are required for the growth of each YBCO monolayer. Given the 7\,Hz laser frequency, this number has been deduced from the time difference between the intensity maxima, respectively for the first YBCO monolayer from the time difference between the start of the growth and the first intensity maximum. It reveals that a significantly smaller number of laser pulses (or a much shorter time) is required for the growth of the first and possibly even the second YBCO monolayer than for the subsequent ones. This unusual effect has been consistently observed for all the YBCO monolayers that have been grown on top of LCMO and even for the ones on the NGO and LSAT substrates. Such an anomalous growth mode of the first YBCO monolayer was also previously noticed for YBCO thin films on SrTiO$_3$ substrates. It was pointed out that depending on the surface termination of the perovskite substrate four different YBCO stacking sequences can be realized. These are CuO-BaO-CuO$_2$-Y-CuO$_2$-BaO (Y-123) or CuO$_2$-Y-CuO$_2$-BaO (Y-112) for the A-site termination and BaO-CuO$_2$-Y-CuO$_2$-BaO (Y-122) or BaO-CuO-BaO-CuO$_2$-Y-CuO$_2$-BaO (Y133) for the B-site termination\cite{Rijnders:2004p14976}. The RHEED data do not allow us to distinguish between them but the STEM data that are discussed below are in favor of Y-122 or Y-112.\\
This raises the question of how a complete monolayer of Y-122 or Y-112 can be grown with a nearly 30\% smaller number of laser pulses than the one required for the subsequent Y-123 monolayers. While the Cu ratio can be accounted for by a missing CuO chain layer, a large discrepancy remains with respect to the Y and Ba ions. One possible explanation is in terms of a substantial interfacial intermixing of Ba and Y with La or Ca. An alternative scenario requires that a significant amount of the deposited material is resputtered or evaporated. The different growth rates of the Y-122 (Y-112) monolayer on LCMO as compared to the subsequent Y-123 monolayers could then by accounted for in terms of different diffusion rates and sticking coefficients of the adatoms.
\subsubsection{Scanning transmission electron microscopy}
\label{STEM}
\begin{figure} [t!]
             \begin{picture}(230,230)
          \put(-5,-5){\rotatebox{0}{\includegraphics[height=230\unitlength]{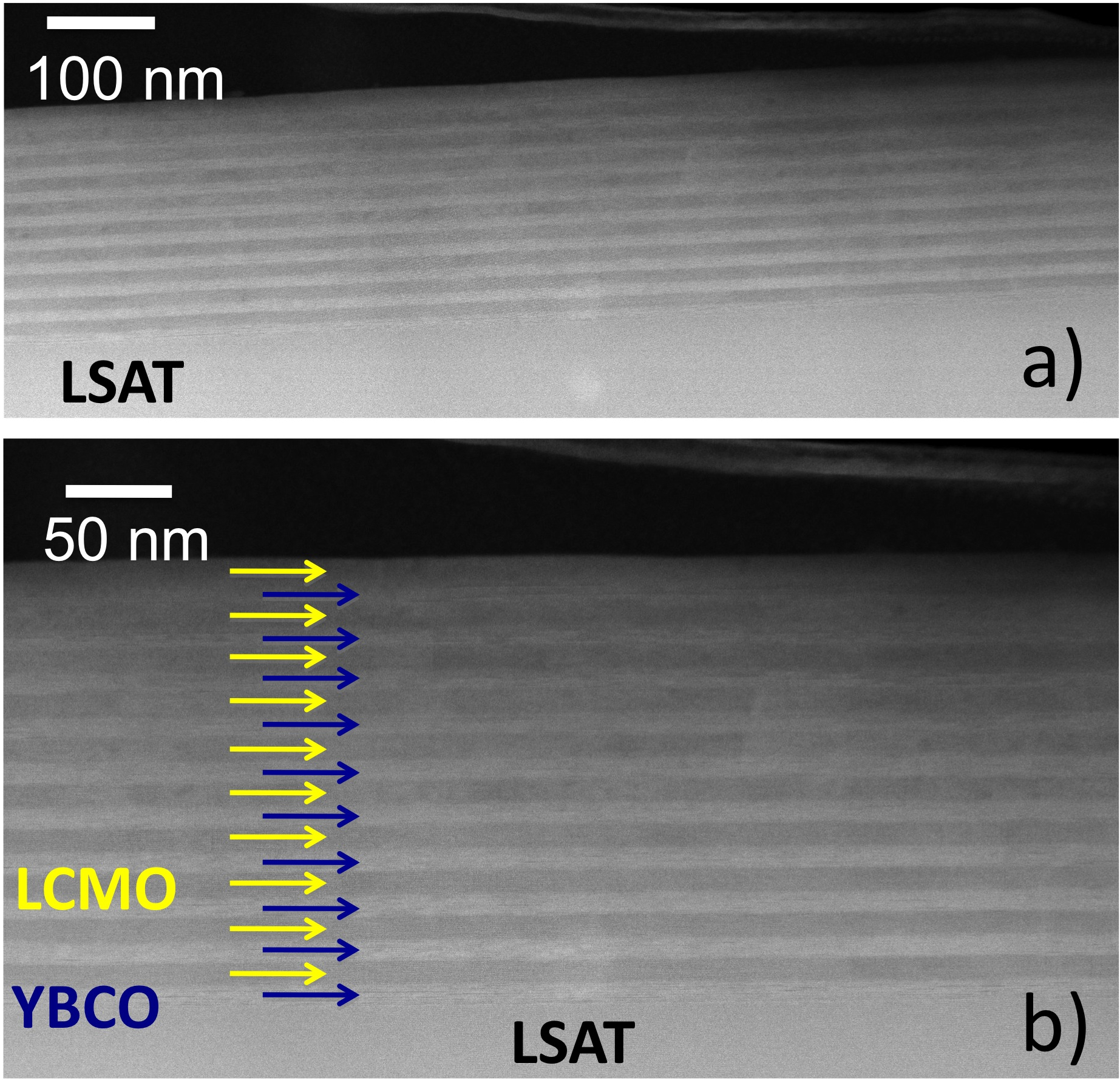}}}
        \end{picture}
        \vspace{0ex}
        \caption{ \label{TEM_1}
     Low magnification STEM images taken on SL-288 with a field of view of about one micrometer (a) and about 500\,nm (b). The individual YBCO and LCMO layers are marked by the blue and yellow arrows, respectively. The layers are very flat and continuous over large distances and there is no sign of any major defects or secondary phases.}
      \end{figure}

\begin{figure} [t!]
             \begin{picture}(270,270)
          \put(0,0){\rotatebox{0}{\includegraphics[height=270\unitlength]{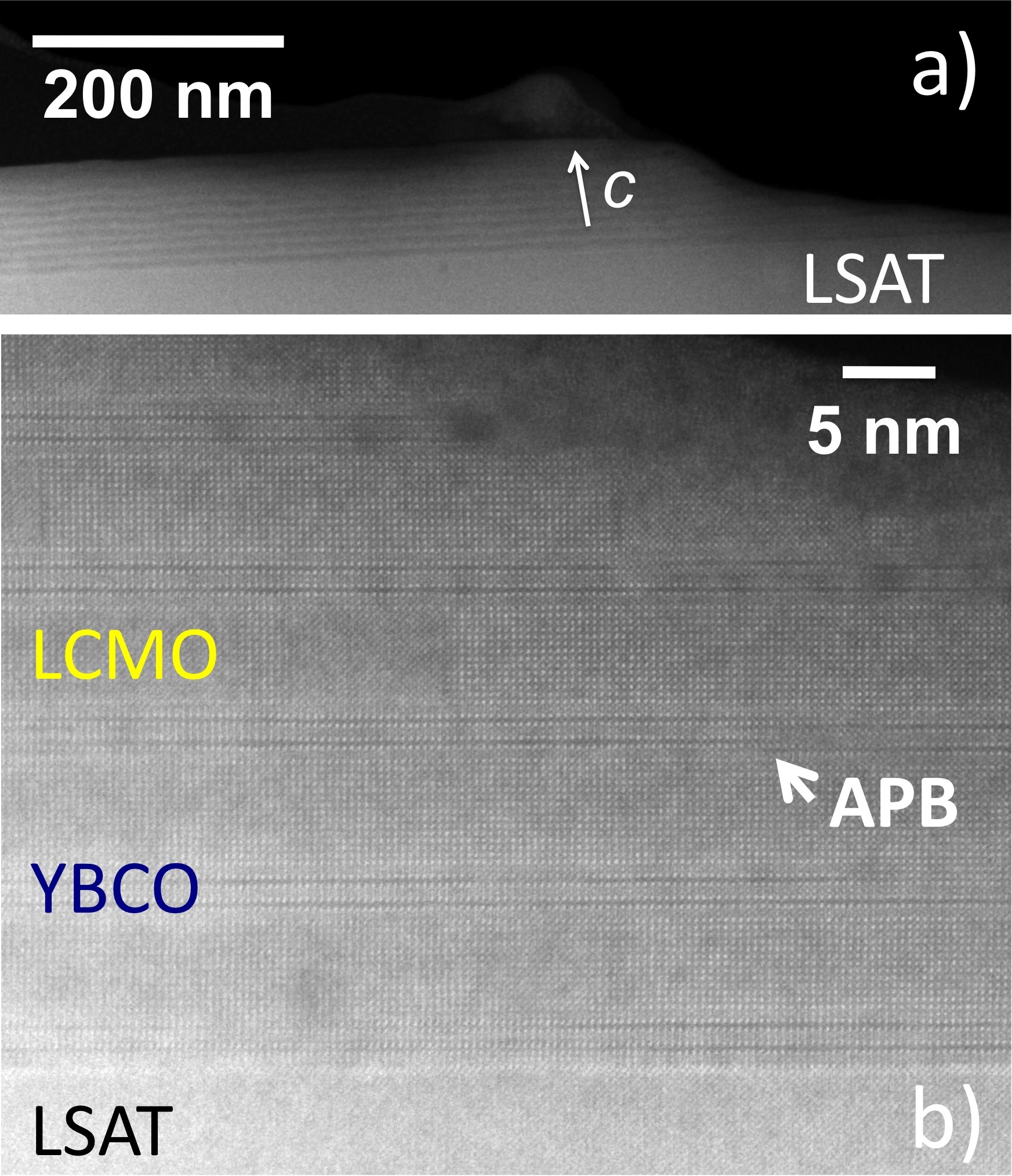}}}
        \end{picture}
        \vspace{0ex}
        \caption{ \label{TEM_3}
     a) Low magnification STEM image of the [YBCO(3)/LCMO(12)]$_{10}$ superlattice (SL-448) grown on a LSAT substrate. b) Corresponding high magnification STEM image showing the individual YBCO (blue) and LCMO (yellow) layers. Each YBCO layer shows two CuO chain planes and three CuO$_2$ bilayers. Only the central CuO$_2$ bilayer has two neighboring CuO chain planes. The high resolution image has been unwarped to remove scan distortions.}
      \end{figure}

The two YBCO/LCMO superlattices on LSAT substrates, [YBCO(8)/LCMO(28)]$_{10}$, SL-288, and [YBCO(3)/LCMO(12)]$_{10}$, SL-448, have been investigated with high resolution STEM and EELS.\\ 
The cross-section of the samples has been scanned with the so-called Z contrast imaging technique where the intensity of every atomic column is roughly proportional to the square of the atomic number, Z. Heavy elements like La, Ba and Y thus give rise to bright columns while the lighter elements like Cu, Mn remain darker, being invisible in the case of O. Fig.~\ref{TEM_1} shows representative low magnification images of SL-288 with a field of view of approximately one micrometer and half a micrometer for $(a)$ and $(b)$, respectively. Such low magnification images show that the individual YBCO and LCMO layers are continuous and flat over long lateral distances and that no major defects or secondary phases are present.\\
      \begin{figure} [h!]
             \begin{picture}(500,500)
          \put(0,0){\rotatebox{0}{\includegraphics[height=500\unitlength]{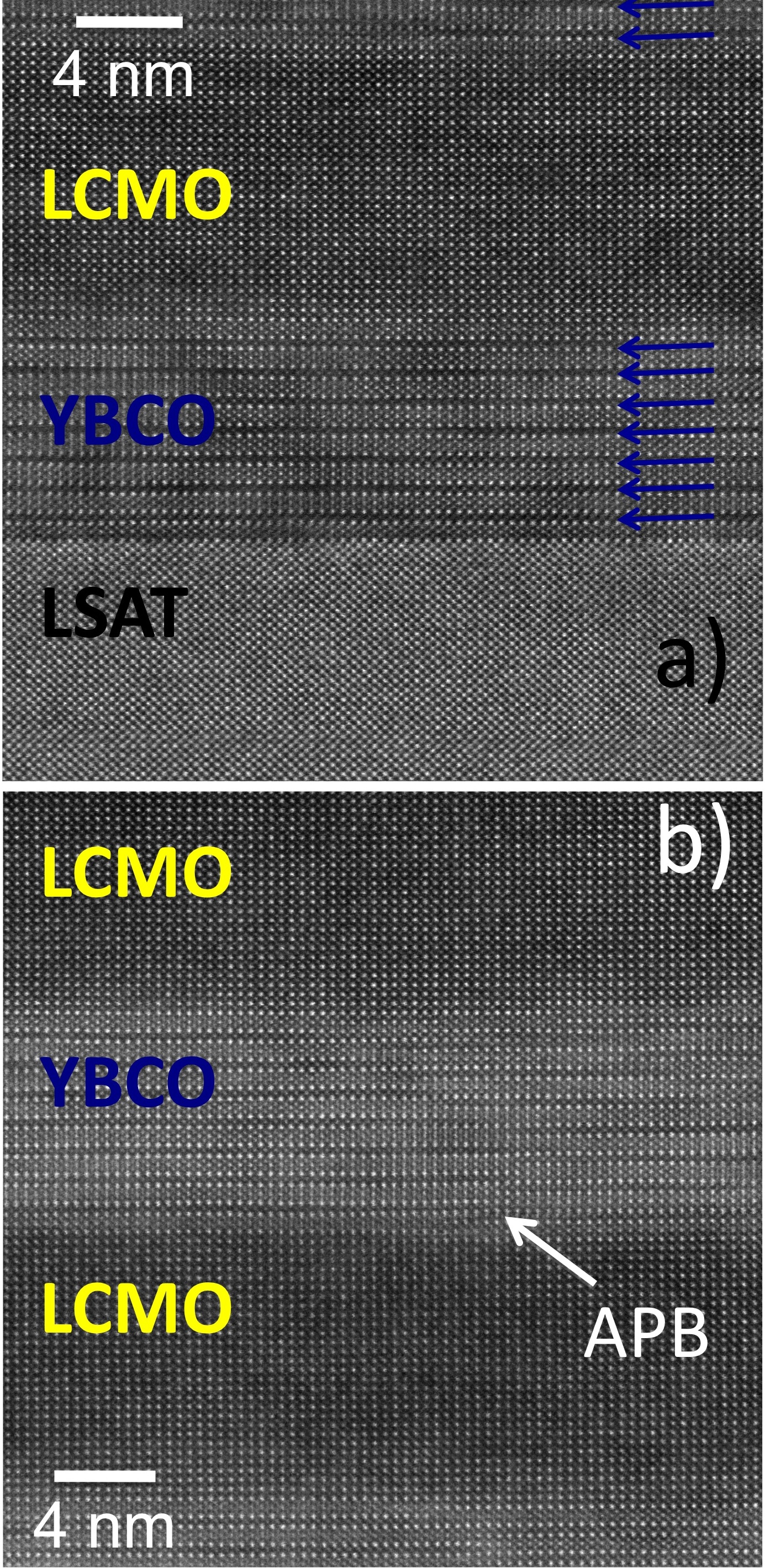}}}
        \end{picture}
        \vspace{0ex}
        \caption{ \label{TEM_2}
     a) High resolution STEM Z-contrast image showing the interface between the LSAT substrate and the first YBCO layer as well as the first YBCO/LCMO bilayers. b) Corresponding STEM image of the YBCO and LCMO layers acquired near the middle of the stacking sequence of SL-288. The images show that the layers are epitaxial and that the interfaces are sharp, coherent and almost defect free. Antiphase boundaries (APB) related to one unit cell high interface steps are occasionally observed, such as the one marked with a white arrow. Images have been unwarped to remove scan distortions.}
      \end{figure}
Fig.~\ref{TEM_2} displays representative high resolution STEM images. Fig.~\ref{TEM_2}~$(a)$ shows a detailed view of the first YBCO/LCMO bilayer on top of the LSAT substrate. A cross-sectional image of some YBCO and LCMO layers near the middle of the SL-288 sample is also shown in Fig.~\ref{TEM_2}~$(b)$. These images highlight that all the layers are epitaxial and that the interfaces are sharp, coherent and nearly free of defects. The only visible defects are occasional antiphase boundaries in the YBCO layer (indicated by the arrow), which originate from single unit cell steps at the interface with the LCMO layer underneath. The individual components of the layered YBCO structure, like the CuO chains and the CuO$_2$ planes, can be clearly identified. The CuO chains appear as darker atomic planes which are framed by bright BaO planes. They are marked by the blue arrows in Fig.~\ref{TEM_2}~$(a)$. The pairs of CuO$_2$ planes (the so-called CuO$_2$ bilayers) are located in the middle of the YBCO unit cell. The Y ions in the center of these CuO$_2$ bilayers are resolved as weaker bright spots. The more isotropic perovskite structure of the cubic LCMO layers is also clearly resolved. Once more the bright spots are the La rich columns while the lighter Mn-O columns show a darker contrast. By counting the number of the atomic layers in the high resolution STEM images of Fig.~\ref{TEM_2} one can easily verify that the YBCO layers have an average thickness of 8 monolayers while the LCMO layers consist of 28 monolayers. This amounts to a total thickness of the YBCO/LCMO bilayer of approximately 200\,\AA\,which agrees rather well with the values that we have deduced from the analysis of the superlattice peaks in the x-ray diffraction and the neutron reflectometry data (shown below).\\
In agreement with the \emph{in-situ} RHEED data, which indicated a different growth dynamics of the first YBCO monolayer on top of LCMO or the LSAT substrate, the STEM images in Figs.~\ref{TEM_2}~$(a)$ and $(b)$ show that this first YBCO monolayer is lacking the CuO chains. The stacking sequence at the LCMO/YBCO interface is (La, Ca)O-MnO$_2$--BaO-CuO$_2$-Y-CuO$_2$-BaO-Y123 (Y-122) or alternatively it may be (La, CaO)--CuO$_2$-Y-CuO$_2$-BaO-Y123 (Y-112). It is difficult to distinguish between these two possibilities since the Ba and the La atoms have a similar contrast in the TEM images. There might also be a mixed (La, Ba, Ca)O layer right at the interface, and possibly even a (Y, La, Ca) layer in the center of the first CuO$_2$ bilayer. In any of these cases, the CuO chains are always missing at the interface monolayers.\\
The subsequent seven YBCO monolayers consist of regular Y-123 units which each contain a layer of CuO chains. As a result there is again a CuO$_2$ bilayer located right at the YBCO/LCMO interface.\\
A corresponding result has been obtained for the [YBCO(3)/LCMO(12)]$_{10}$ superlattice (SL-448) for which the low- and high magnification TEM images are shown in Fig.~\ref{TEM_3}. Again all the layers of the superlattice are flat and coherent over a lateral distance on the order of a micrometer (the STEM experiment field of view). And once more for both the YBCO/LCMO and the LCMO/YBCO interfaces there is a CuO$_2$ bilayer located right at the interface, i.e. each YBCO layer contains three CuO$_2$ bilayers but only two layers of CuO chains. The interfacial configuration with a CuO$_2$ bilayer that is connected to a MnO$_2$ layer thus appears to be most favorable irrespective of whether it fits in with the regular stacking sequence of Y-123 (for the YBCO/LCMO interface) or whether it requires a modified Y-112 or Y-122 stacking sequence (for the LCMO/YBCO interface).
The same stacking sequence with a missing CuO chain layer at the LCMO/YBCO and the YBCO/LCMO interfaces was previously observed for superlattices on SrTiO$_3$ substrates that were grown with a high oxygen pressure sputtering technique\cite{Varela:2003p17134}. Nevertheless, a different result was reported by Zhang and coworkers who observed an asymmetric termination with a CuO chain layer at one of the interfaces and a CuO$_2$ bilayer at the other one for a PLD grown trilayer structure of YBCO/LCMO/YBCO on SrTiO$_3$\cite{Zhang:2009p14937}. This is especially surprising since the PLD growth was performed under similar growth conditions as for our present samples. The reason for these differences in the interfacial termination is unknown to us. One possible factor may be the thicknesses of the individual YBCO layers which was about 50\,nm in Ref.~\cite{Zhang:2009p14937} as compared to 10\,nm or less in our present study and the one of Ref.~\cite{Varela:2003p17134}. It was indeed previously found that a coherent layer by layer growth mode of YBCO can only be maintained up to a film thickness of about 15\,nm \cite{Dam:2002p17856}. Another important factor may be the surface termination of the substrate which has not been well controlled in our present and these previous studies. Clearly, more systematic work in this direction will be required in order to identify the parameters that determine the interfacial layer sequences of these YBCO/LCMO superlattices.\\
      \begin{figure} [b!]
             \begin{picture}(185,185)
          \put(-28,0){\rotatebox{0}{\includegraphics[height=185\unitlength]{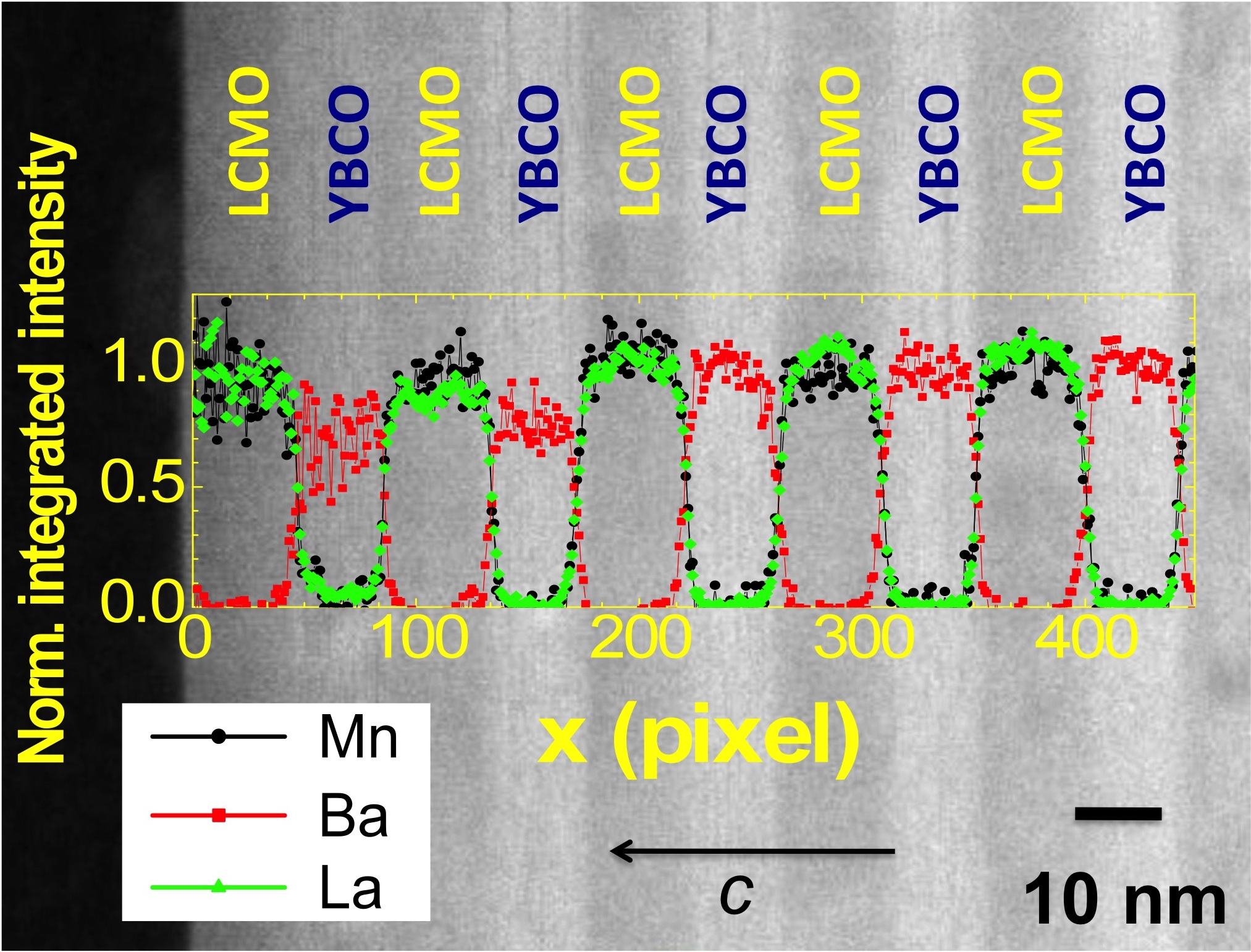}}}
        \end{picture}
        \vspace{0ex}
        \caption{ \label{TEM_4}
   Analysis of the chemical composition with electron energy loss spectroscopy (EELS) for SL-288. The normalized integrated intensities along the growth direction are plotted for the  Mn L$_{2,3}$ (black), Ba M$_{4,5}$ (red), and La M$_{4,5}$ (green) edges. A low magnification STEM image of the sample is shown in the background on a matching scale. The EELS data confirm that the interfaces are sharp and exhibit no indication for major chemical interdiffusion. The broadening of the edges of the profiles at the interfaces is very likely an artifact associated to broadening of the electron beam due to dechanneling.}
      \end{figure}
We have also investigated the chemical composition of SL-288 with spatially resolved electron energy loss spectroscopy (EELS). Fig.~\ref{TEM_4} shows how the normalized EELS intensities for the Mn (black), La (green) and Ba (red) edges evolve along the growth direction of the superlattice. This element specific analysis confirms that the interfaces are sharp and that no significant chemical interdiffusion occurs. The broadening of the interfacial steps in the profiles is most likely due to the finite width of the electron beam because of dechanneling. The thickness of specimens prepared with these methods is typically of 0.2-0.5 inelastic mean free paths (i.e. a few tens of nm), so significant beam broadening is expected, which limits the spatial sensitivity of this technique. Accordingly, we cannot exclude that some chemical interdiffusion occurs right between the very first YBCO and LCMO monolayers right at interface.\\
      \begin{figure} [h!]
             \begin{picture}(335,335)
          \put(-5,0){\rotatebox{0}{\includegraphics[height=335\unitlength]{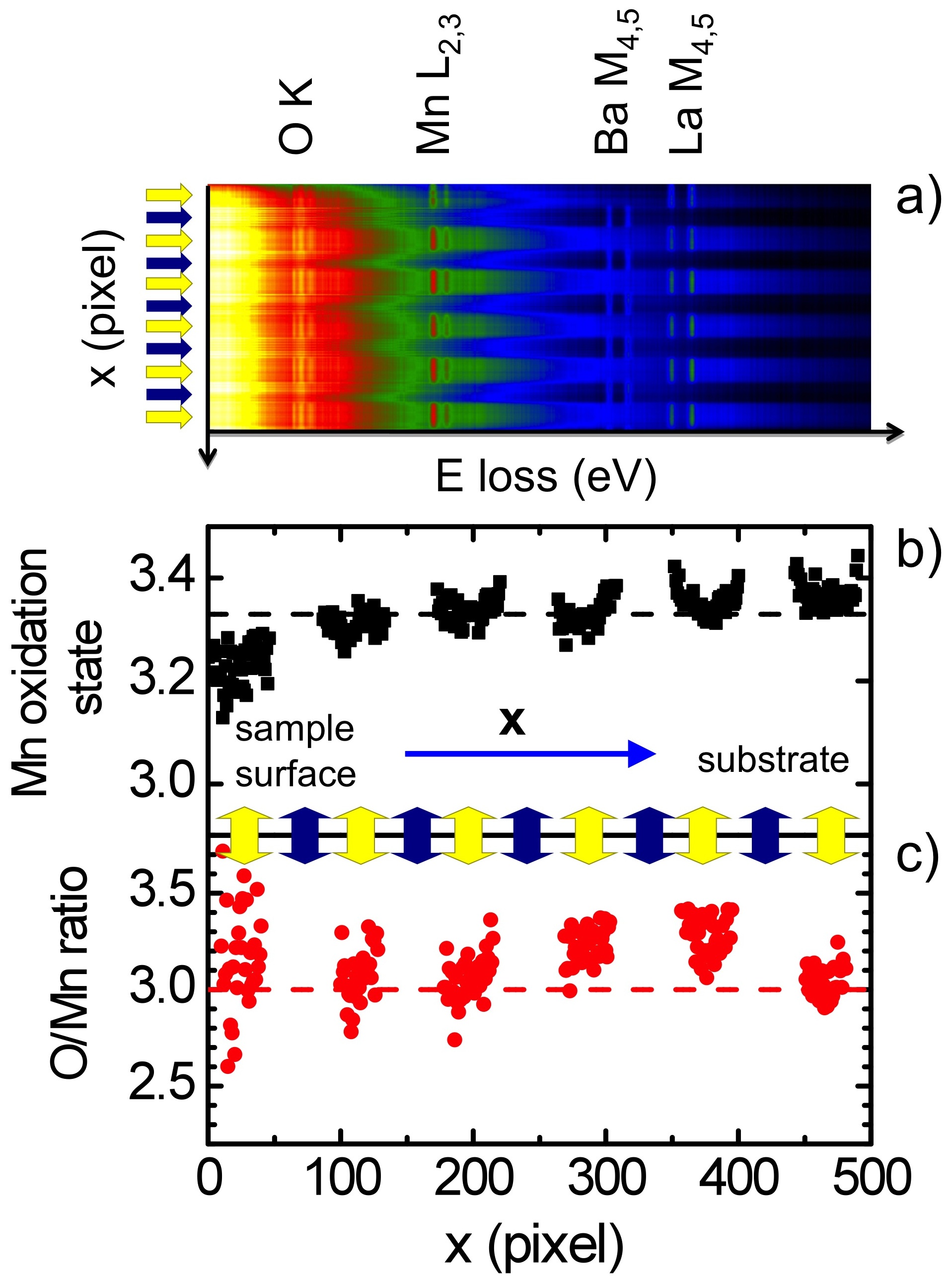}}}
        \end{picture}
        \vspace{0ex}
        \caption{ \label{TEM_5}
   EELS line scan obtained by scanning the electron beam along the stacking direction of the superlattice starting from the surface and going towards the substrate. The spectra show the O K, Mn L$_{2,3}$, Ba M$_{4,5}$ and La M$_{4,5}$ edges. b) Mn oxidation state as calculated from the separation between the pre- and main peaks of the O K edge. c) O/Mn ratio obtained from the chemical quantification analisys available in the Gatan Digital Micrograph software. Blue (yellow) arrows marke the positions of the YBCO (LCMO) layers along the scan.}
      \end{figure}
Fig.~\ref{TEM_5}~$(a)$ shows a color plot of the intensity variation of the EELS signal at the O-K, Mn-L$_{2,3}$, Ba-M$_{4,5}$, and La-M$_{4,5}$ edges across the individual layers of SL-288. Fig.~\ref{TEM_5}~$(b)$ shows the oxidation state of the Mn ions that has been calculated from the separation in energy between the prepeak and the main peak of the O-K edge. The obtained value of about +3.3 is very close to the nominally expected one for La$_{2/3}$Ca$_{1/3}$MnO$_3$ (marked with a horizontal dashed line)\cite{Varela:2009p19397}. Note that the slight increase of the Mn oxidation state towards the interfaces is an artifact that arises due to oversampling associated to the finite width of the electron beam which partially probes the O-K edge of YBCO near the interfaces. A significant feature is the reduced Mn oxidation state near +3.2 measured at the topmost LCMO layer. This observation could be caused by a partial degradation or deoxygenation of this top LCMO layer, which is exposed to the ambient and thus also to some moisture. Such degradation of the first few monolayers on the surfaces of LCMO thin films has also been observed with X-ray photoelectron spectroscopy\cite{Schwier2011}. It also explains that the topmost LCMO layer has a significantly reduced ferromagnetic magnetization density as is shown with polarised neutron reflectivity measurements (see paragraph~\ref{Magnetic_prop}). Finally, Fig.~\ref{TEM_5}~$(c)$ shows that the average value of the O/Mn ratio, obtained from the chemical quantification analisys routine of Digital Micrograph, is close to the nominal value of 3 (marked with a horizontal dashed line). While the error bars for this quantification method can be relatively large, it appears again that the O/Mn ratio of the top layer is somewhat decreased as compared to the inner LCMO layers of the SL. This finding is in agreement with the increasing manganese valence observed in the layers near the substrate.
\subsubsection{X-ray diffraction and neutron reflectometry}
\label{XRD_NR}
\begin{figure} [h!]
       \begin{picture}(350,350)
          \put(-5,0){\rotatebox{0}{\includegraphics[height=350\unitlength]{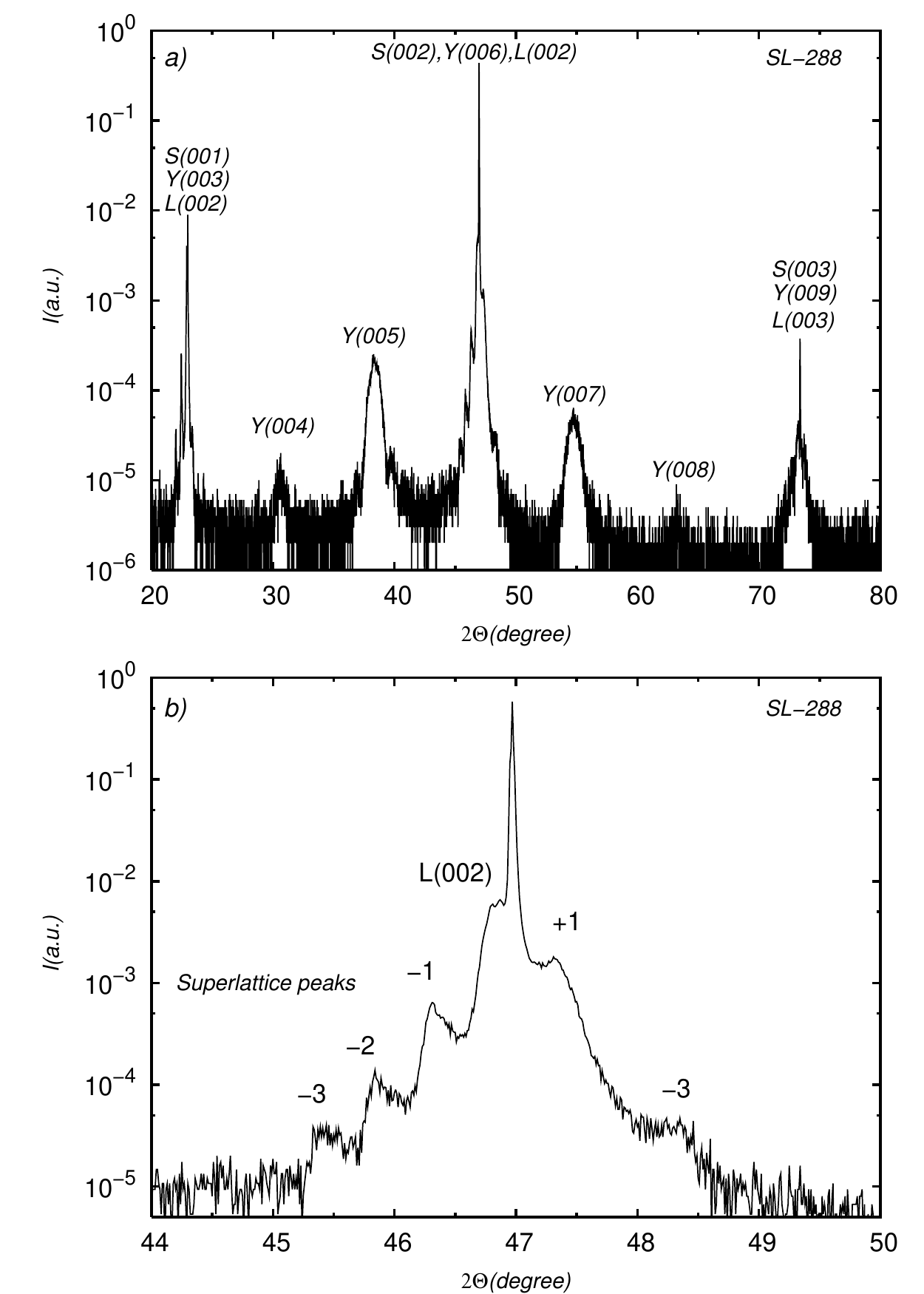}}}
        \end{picture}
        \vspace{0ex}
        \caption{ \label{XRD}
       a) X-ray diffraction curve ($\Theta$-$2\Theta$ scan) of [YBCO(8)/LCMO(28)]$_{10}$ SL-288 that confirms the epitaxial growth and the high structural quality of the superlattice. The (00$l$) peaks of YBCO, LCMO and the LSAT substrate are labeled as Y, L, and S, respectively. b) Magnification around the Y(006), L(002), and S(002) peaks. The pronounced satellite peaks testify for the high structural quality of the superlattice.}
      \end{figure}

\begin{figure} [b!]
       \begin{picture}(175,175)
          \put(-35,0){\rotatebox{0}{\includegraphics[height=175\unitlength]{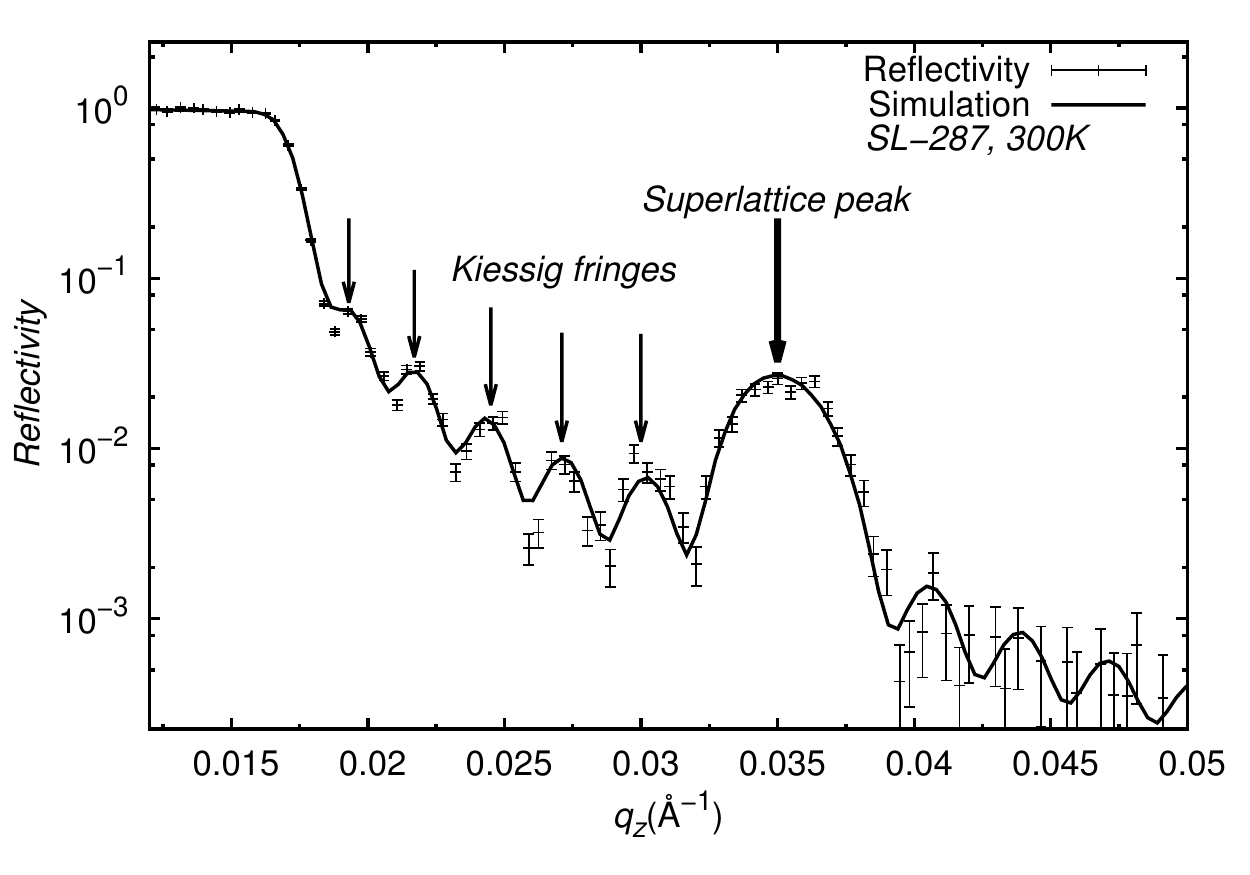}}}
        \end{picture}
        \vspace{0ex}
        \caption{ \label{NR}
        Unpolarized, specular neutron reflectivity curves as obtained at room temperature for superlattice [YBCO(8)/LCMO(28)]$_{10}$ (SL-287). The symbols show the experimental data, the solid lines the simulation with the supefit program. The simulation parameters are listed in table~\ref{nvalue_table}}
      \end{figure}

The structural quality of SL-288 has been investigated further by x-ray diffraction measurements. A corresponding $\Theta$-$2\Theta$ scan is shown in Fig.~\ref{XRD}.  The sharp (00$l$) reflections of YBCO and LCMO in Fig.~\ref{XRD}~$(a)$ confirm that all the layers are fully c-axis oriented and epitaxial. Fig.~\ref{XRD}~$(b)$ shows a magnification of one of the peaks which reveals a series of pronounced satellite peaks. These correspond to an interference pattern that arises when the x-ray beams which are reflected from the individual atomic planes in the YBCO or LCMO layers remain phase coherent. The mere observation of these pronounced satellite peaks testifies for the high perfection and the crystalline quality of this superlattice. The spacing between these peaks yields an estimate for the thickness of the YBCO/LCMO bilayers of 200.4\,\AA. This value agrees well with the ones that have been deduced from the STEM data (see paragraph~\ref{STEM}) as well as from the neutron reflectometry data on the nominally identical superlattice SL-287, as is shown below.\\
Additional information about the structural parameters and the quality of the superlattices has been obtained from specular neutron reflectometry measurements. The neutron measurements probe the entire area of these 10$\times$10\,mm$^2$ sized superlattices and therefore provide complementary information with respect to the STEM data which probe the structural properties on the local micrometer scale and also the x-ray diffraction data which probe only the central part of the film (on the lateral scale of millimeters).  Representative neutron reflectivity curves for the superlattices and [YBCO(8)/LCMO(28)]$_{10}$ (SL-287) are displayed in Fig.~\ref{NR}. The SL-287 has been grown under identical conditions and is nominally identical to SL-288 which has been investigated with x-ray diffraction and STEM.
The reflectivity profile contains clear Kiessig fringes that are  marked by the arrows. These arise from the interference between the neutron waves that are reflected from the top and the bottom layers of the superlattice, i.e. from the ambient/superlattice and the superlattice/substrate interfaces. The presence of such pronounced Kiessig fringes confirms that the superlattice has a very uniform thickness and a fairly low surface roughness. From their period one can readily deduce the total thickness of the superlattice. The reflectivity curves also contain strong superlattice peaks that arise from the interference between the waves that are reflected from all the YBCO/LCMO and LCMO/YBCO interfaces. The position, height and width of these superlattice peaks contains the information about the thickness of the YBCO/LCMO bilayers, the contrast in the scattering density of the YBCO and LCMO layers and their uniformity across the entire 10$\times$10\,mm$^2$ sized film. The solid line in Fig.~\ref{NR} shows the simulation which has been performed with the supefit program\cite{Ruhm:1999p17046}. The simulation parameters are listed in table~\ref{nvalue_table}. The thicknesses of the YBCO and LCMO layers as deduced from the neutron reflectometry measurements of SL-287 agree reasonably well with the estimates that have been obtained from the x-ray diffraction and STEM data for the nominally identical SL-288. The root mean square (rms) roughness of the LCMO and YBCO layers is rather small, i.e. the absolute values are on the length scale of unit cell height steps of LCMO and YBCO.\\
The \emph{in-situ} RHEED, STEM, x-ray diffraction and neutron reflectometry data provide complementary information about the structural properties of our YBCO/LCMO superlattices. The combined information suggests that the PLD growth of these superlattices has been controlled on the scale of single monolayers along the vertical direction while a coherent and homogeneous layer growth has been maintained on the millimeter scale. This combined study has also provided clear evidence that growth dynamics and the resulting stacking sequence of the atomic layers at the LCMO/YBCO interfaces is rather unusual, i.e. that the first YBCO monolayer is always lacking its CuO chain layer. This study should motivate further experimental work where for example the surface termination of the substrate is well determined and/or the influence of the variation of the thickness of the individual YBCO and LCMO layers are studied more systematically.
\subsection{Electromagnetic properties}
\label{EM_prop}
\subsubsection{Electronic properties}
\label{Electronic_prop}
\begin{figure} [b!]
       \begin{picture}(175,175)
          \put(-35,0){\rotatebox{0}{\includegraphics[height=175\unitlength]{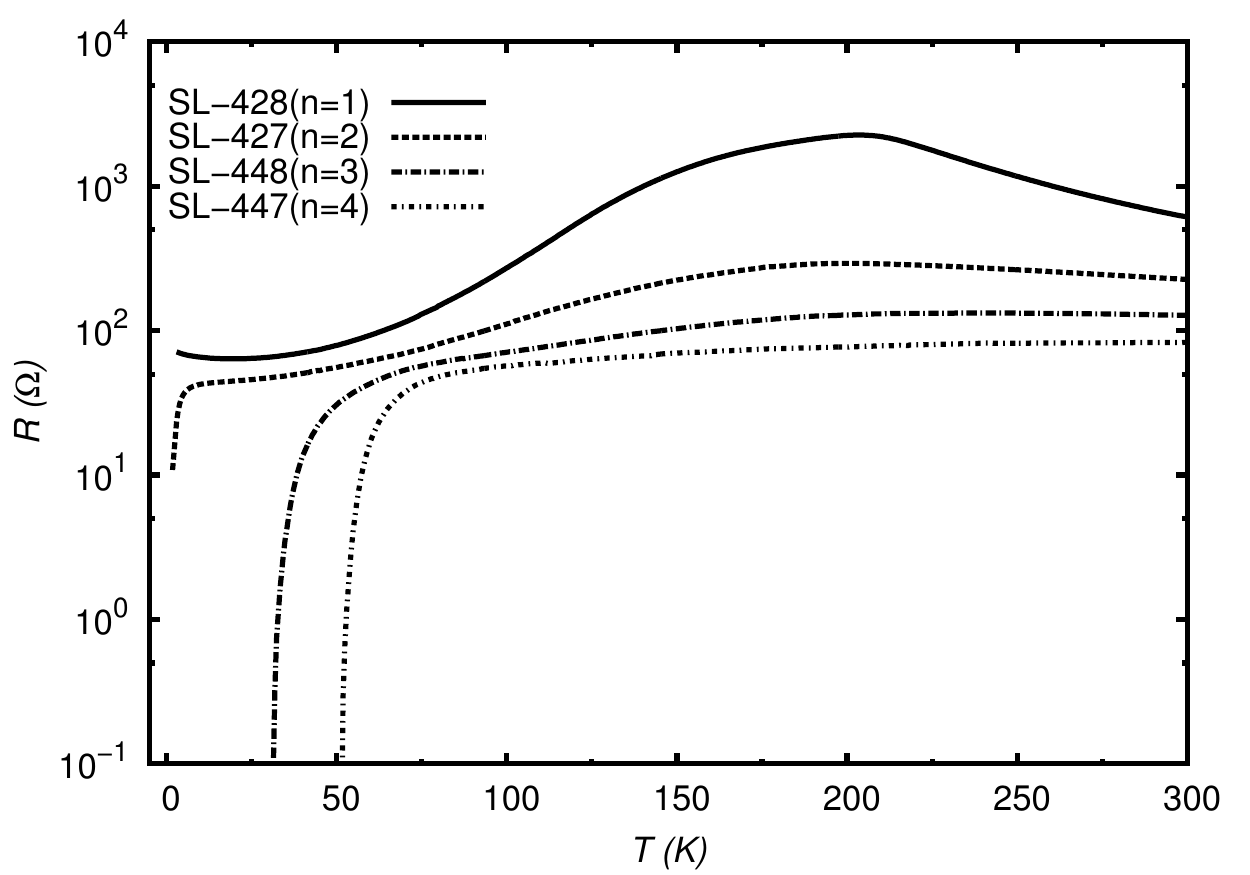}}}
        \end{picture}
        \vspace{0ex}
        \caption{Resistance versus temperature for [YBCO(n)/LCMO(m)]$_{10}$ superlattices with n=1-4 and m=12.}
         \label{Electric_thinfilm} 
      \end{figure}

\begin{figure} [h!]
       \begin{picture}(175,175)
          \put(-35,0){\rotatebox{0}{\includegraphics[height=175\unitlength]{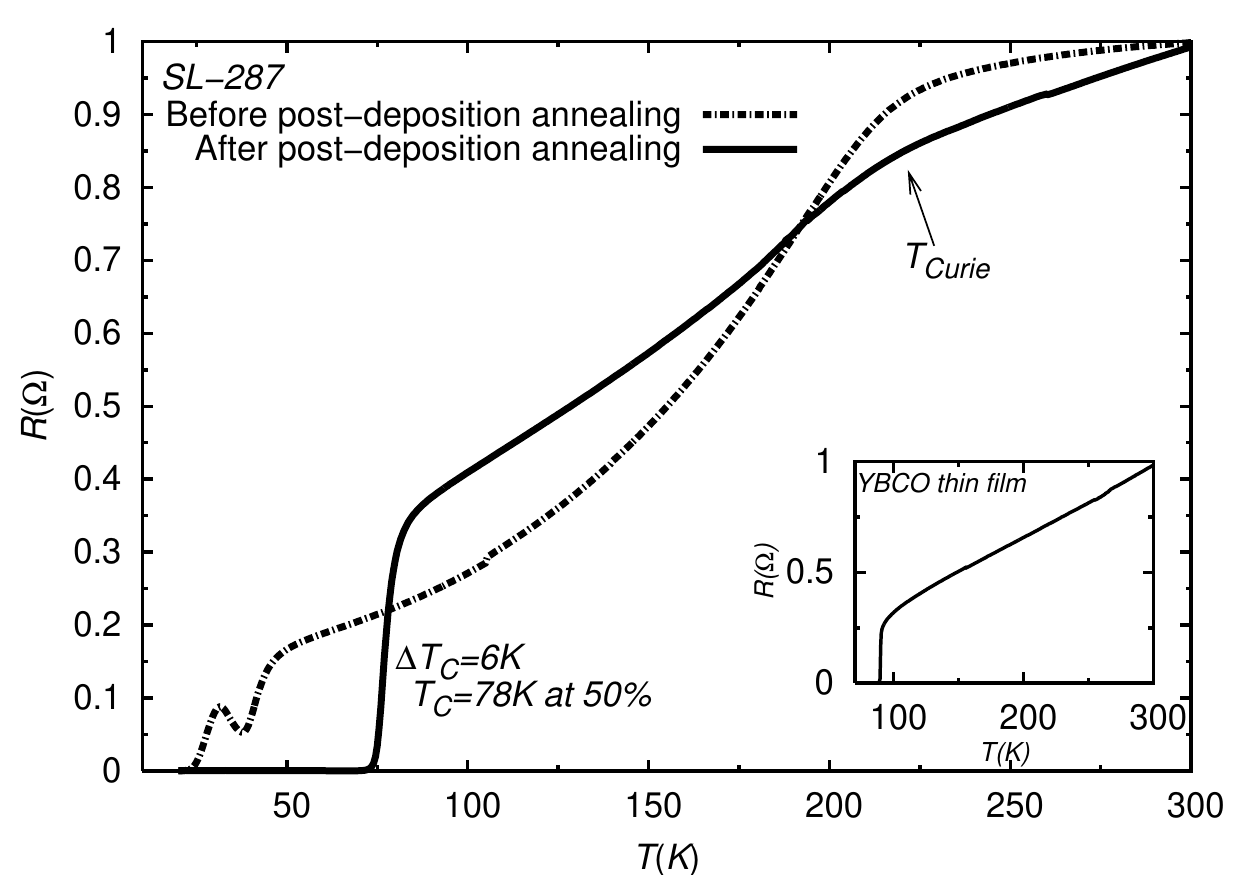}}}
        \end{picture}
        \vspace{0ex}
        \caption{Resistance of SL-287 before (dash-dotted line) and after (solid line) the \emph{ex-situ} annealing treatment.  Inset: Corresponding data before the \emph{ex-situ} annealing for a single 100\,nm thick YBCO thin film. No significant difference was observed here after the \emph{ex-situ} annealing treatment.}
         \label{Electric_SL} 
      \end{figure}

The STEM images as shown above establish that the CuO$_2$ bilayers that are situated right at the LCMO/YBCO or YBCO/LCMO interfaces have only one neighboring CuO chain layer since on the side towards the interface they are connected to a MnO$_2$ layer. As was already previously pointed out by Varela and coworkers\cite{Varela:2003p17134}, this has important consequences for the hole doping state of these CuO$_2$ bilayers which is determined by the transfer of electrons between the CuO$_2$ bilayers and the neighboring CuO chains. As the oxygen content of the CuO chains is increased, there are successively more holes created in the CuO$_2$ bilayers, the optimal hole doping state with a $T_\mathrm{C}$ value of about 90\,K is achieved when the CuO chains are almost fully oxygenated. A decrease in the oxygen content of the CuO chain first results in a decrease of $T_\mathrm{C}$ in the underdoped state and finally in a metal-to-insulator transition as the CuO$_2$ bilayers become undoped and a Mott-Hubbard gap develops due to the strong correlations of the electrons in the half filled Cu-$3d_\mathrm{x^2-y^2}$ band. Accordingly, it can be expected that the hole doping state of the interfacial CuO$_2$ bilayers is significantly reduced, since they are lacking one of their neighbouring CuO chains and thus half of their charge reservoir. In addition, the electronic properties of the interfacial CuO$_2$ bilayers might be affected by a charge transfer across the interface, by an orbital reconstruction of the relevant interfacial states as reported in Ref.~\cite{CHAKHALIAN:2007p3736}, or by a chemical intermixing of La and Ba on the interfacial BaO layer or of Y and Ca on the Y layer of the interfacial CuO$_2$ bilayer. The interfacial CuO$_2$ bilayers therefore are likely less conducting, they might even be entirely insulating and act as barriers which block the electronic proximity coupling between the superconducting YBCO and the ferromagnetic LCMO layers.\\
In the following we present the electric transport properties of a series of [YBCO(n)/LCMO(m)]$_{10}$ superlattices with n=1-4 and m=12 which provide evidence that the interfacial CuO$_2$ bilayers are strongly underdoped, yet they remain conducting and even exhibit the onset of a superconducting transition at very low temperature. The corresponding temperature dependent resistance curves are shown in Fig.~\ref{Electric_thinfilm}. For the sample with n=1 there is no trace of a superconducting transition down to the lowest measured temperature of 2\,K as observed in previous reports\cite{Sefrioui:2002p19006}. The electric transport is dominated by the LCMO layers as is evident from the pronounced maximum in the resistance curve around 200-220\,K. The latter is a characteristic feature of the concomitant ferromagnetic-to-paramagnetic and metal-to-insulator transition of LCMO which is at the heart of the so-called colossal magnetoresistance (CMR) effect. This behavior confirms the expected trend that the isolated CuO$_2$ bilayers without any neighboring CuO chains remain insulating. However, already for the n=2 sample for which the two CuO$_2$ bilayers are sharing a single layer of CuO chains, there are significant changes in the resistance which suggest that these YBCO layers are conducting. With respect to the n=1 sample, the value of the conductance is significantly reduced. In particular, the peak around the Curie temperature of LCMO is strongly suppressed. There is even a sharp drop in the resistance at low temperature which provides evidence for the onset of superconductivity below ${T_\mathrm{C}^\mathrm{onset}}$\,$\approx$\,8\,K. This highlights that the two CuO$_2$ bilayers which share only one CuO chain layers as a charge reservoir are conducting and at very low temperature they even reveal the onset superconductivity. For the samples with n=3 and n=4 the resistance decreases further and the superconducting transition temperature rises rather rapidly to values of $T_\mathrm{C}$(R$\rightarrow$0)\,$\approx$\,34\,K and 53\,K, respectively. A clear downturn in the resistance due to the onset of superconductivity is observed at ${T_\mathrm{C}^\mathrm{onset}}$\,$\approx$\,75K and 80K, respectively.\\
In this context we emphasize that the oxygenation of the YBCO layers of our superlattices turned out to be a surprisingly slow process. The oxygen annealing treatment that was performed \emph{in-situ} right after the PLD growth of the thin films was not sufficient to obtain a fully oxygenated state of the YBCO layers. This state required an additional annealing treatment which was performed \emph{ex-situ} in a separate furnace with a gas flow of pure oxygen (100\,ml/min) for 12 hours at 485\,$^{\circ}$C with subsequent slow cooling to room temperature. While for single thin films of YBCO and also for YBCO/LCMO superlattices grown on SrTiO$_3$ substrates the \emph{in-situ} annealing procedure was sufficient to obtain fully oxygenated films with the highest $T_\mathrm{C}$ values, for our present YBCO/LCMO superlattices on NGO and LSAT substrates this resulted in very broad corresponding superconducting transitions with significantly reduced $T_\mathrm{C}$ values.\\
As an example, Fig.~\ref{Electric_SL} shows the temperature dependence of the resistance of SL-287 before and after the \emph{ex-situ} annealing treatment by the dash-dotted and the solid lines, respectively. In both cases the resistance exhibits a noticeable kink around 225\,K which indicates that the ferromagnetic transition of the LCMO layers with $T_\mathrm{Curie}$\,$\approx$\,225\,K is not significantly affected by this additional annealing. However, the superconducting transition temperature of the YBCO layers is obviously strongly affected by this additional annealing step. Before the \emph{ex-situ} annealing, the transition occurs at a rather low temperature and it is very broad with an onset around 41\,K and a true zero resistance below 27\,K. In addition, a pronounced minimum-maximum structure occurs in the intermediate temperature range that is suggestive of a spatial inhomogeneity of the superconducting order parameter which leads to interference effects that arise from the mismatch of the phase of the order parameter in different regions of the sample\cite{Mosqueira:1994p18550}. After the \emph{ex-situ} annealing step in flowing oxygen, this unusual structure is absent and a relatively sharp superconducting transition is observed with $T_\mathrm{C}$(R$\rightarrow$0)\,$\approx$\,75K and ${T_\mathrm{C}^\mathrm{onset}}$\,$\approx$\,80K. In fact, the transition is nearly as sharp as the one of a 100\,nm thick single film of YBCO for which the resistance curve is shown in the inset of Fig.~\ref{Electric_SL}. 
These observations suggest that the oxygen partial pressure during the PLD growth and the subsequent \emph{in-situ} annealing, while they enable one to obtain stoichiometric LCMO layers, are not sufficient to achieve a full oxygenation of the YBCO layers. We believe that the apparently very slow process of the equilibration of the oxygen content of the YBCO layers in these superlattices is correlated with their high crystalline perfection. More specifically we assume that it is related to their low density of extended defects, like grain boundaries or screw dislocations, which can act as shortcuts for the oxygen diffusion paths since they may significantly enhance the oxygen mobility along the vertical direction of the superlattices and thus reduce the lateral diffusion length. In YBCO it is indeed well established that the oxygen mobility along the c-axis direction (perpendicular to the CuO$_2$ bilayers and the CuO chains) is extremely small and that very long annealing times are required for YBCO single crystals where the oxygen thus needs to diffuse over long lateral distances along the direction of the CuO chains. In agreement with this conjecture, we have found that for superlattices with a lower structural quality and higher defect density, the \emph{ex-situ} annealing treatment is not required (or at least it can be much shorter) to achieve sharp and relatively high superconducting transitions. In contrast, for our YBCO/LCMO superlattice with n=1 to 4 the \emph{ex-situ} annealing procedure was absolutely essential, i.e. no signature of a superconducting transition was observed without this additional treatment.

\subsubsection{Magnetic properties}
\label{Magnetic_prop}
In the previous paragraph we have already provided evidence, from the temperature dependence of the resistance data, that the ferromagnetic transition in our YBCO/LCMO superlattices (for n=1,2,3,4, and 10 and m=12 and 28) generally occurs well above 200\,K. It was also shown that contrary to the superconducting transition temperature and the metallic properties of the YBCO layers, this magnetic transition was not strongly affected by the \emph{ex-situ} annealing treatment. In the following we present some dc magnetization and polarized neutron reflectometry (PNR) measurements which establish that the ferromagnetic order within the LCMO layers is indeed well established and involves a sizeable average magnetic moment of about 2.7\,$\mu_\mathrm{B}$ per Mn ion that compares rather well to the one of 3.6\,$\mu_\mathrm{B}$ in bulk LCMO single crystals\cite{Wollan55,Salamon01}.\\
      \begin{figure} [h!]
       \begin{picture}(350,350)
          \put(0,0){\rotatebox{0}{\includegraphics[height=350\unitlength]{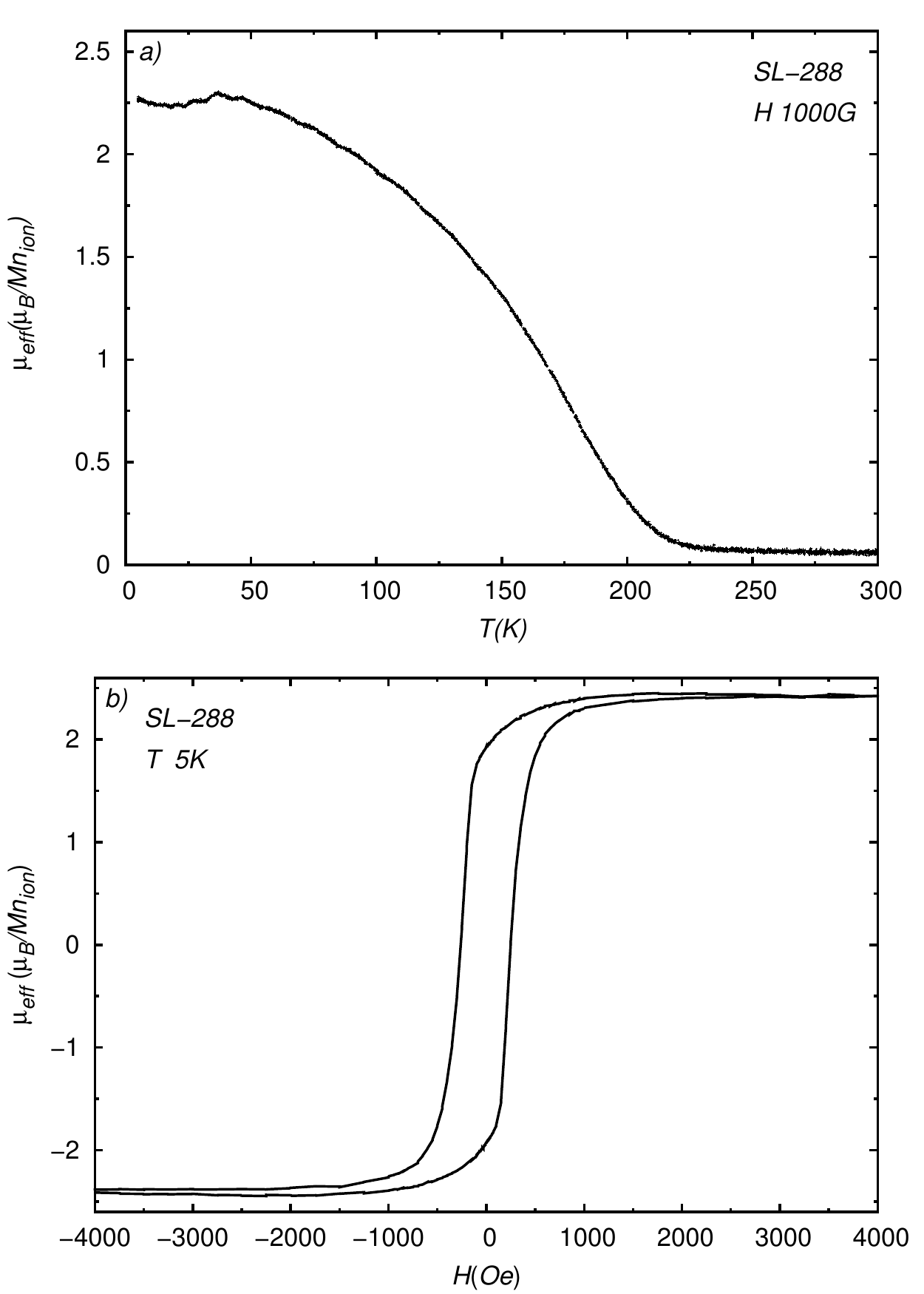}}}
        \end{picture}
        \vspace{0ex}
        \caption{(a) Magnetic moment per Mn ion (in units of Bohr magnetons) as measured upon field-cooling in an external magnetic field of 1000\,Oe that was applied parallel to the layers of the superlattice SL-288. (b) M-H magnetization loop measured at T=5\,K}
         \label{Mag_SL_VSM} 
      \end{figure}
\begin{table}
\hfill
  \begin{center}
    \begin{tabular}{|c|c|c|}
      \hline
$Temparture$	& $300\,\mathrm{K}$	& $10\,\mathrm{K}$ \cr
      \hline
$\rho^\mathrm{NGO}$($\times10^{-6}\mathrm{\AA}^{-2}$)			& 5.644			& 5.644 \cr
      \hline
$\rho^\mathrm{YBCO}$($\times10^{-6}\mathrm{\AA}^{-2}$)				&4.572			&4.572 \cr
      \hline
$\rho^\mathrm{LCMO}$($\times10^{-6}\mathrm{\AA}^{-2}$)		&3.543				&3.543 \cr
      \hline
$\rho^\mathrm{LCMO(top)}$($\times10^{-6}\mathrm{\AA}^{-2}$)			& 3.074				& 3.068 \cr
      \hline         
$d^\mathrm{YBCO(bottom)}$($\mathrm{\AA}$)			& 93.6				& 93.3 \cr
      \hline         
$\sigma^\mathrm{YBCO(bottom)}$($\mathrm{\AA}$)		& 8.5			& 9.0 \cr
      \hline         
$d^\mathrm{YBCO}$($\mathrm{\AA}$)			& 93.6				& 93.3 \cr
      \hline         
$\sigma^\mathrm{YBCO}$($\mathrm{\AA}$)		& 8.5			& 9.0 \cr
     \hline
$d^\mathrm{LCMO}$($\mathrm{\AA}$)			& 100.3				& 100.0 \cr
      \hline         
$\sigma^\mathrm{LCMO}$($\mathrm{\AA}$)		& 8.0			& 8.0 \cr
      \hline
 $d^\mathrm{LCMO(top)}$($\mathrm{\AA}$)			& 93.6				& 93.3 \cr
      \hline         
$\sigma^\mathrm{LCMO(top)}$($\mathrm{\AA}$)		& 15.3			& 15.3 \cr
      \hline 
      $B$ ($\mu_\mathrm{B}/\mathrm{Mn}$)			& 0				& 2.7 \cr
      \hline         
$B^{top layer}$ ($\mu_\mathrm{B}/\mathrm{Mn}$)		& 0			& 2.3 \cr
      \hline          
    \end{tabular}
    \caption{\label{nvalue_table}
  Parameters for the simulation of the neutron reflectivity spectra at 300\,K and 10\,K as shown in Fig.~\ref{NR} and Fig.~\ref{Mag_SL_PNR}.
    }
  \end{center}
\end{table}
 Fig.~\ref{Mag_SL_VSM} shows the magnetization data of SL-288 rescaled in terms of the magnetic moment per Mn ion of the LCMO layers. Fig.~\ref{Mag_SL_VSM}~$(a)$ shows the temperature dependence of the field cooled measurement where an external magnetic field of 1000\,Oe was applied parallel to the layers of the superlattice. It reveals the onset of a spontaneous ferromagnetic order below $T_\mathrm{Curie}$\,$\approx$\,225\,K. This value agrees well with the estimate from the temperature dependence of the resistance data that are shown in Fig.~\ref{Electric_SL}. Below $T_\mathrm{Curie}$ the magnetic signal increases rather rapidly and below 100\,K it reaches values in excess of 2\,$\mu_\mathrm{B}$ per Mn ion. An anomalous decrease below about 30\,K is most likely related to a superconducting screening effect. This screening evidently sets in well below the superconducting transition temperature at $T_\mathrm{C}$\,$\approx$\,75\,K. A similar effect has been previously observed by several authors\cite{habermeier:2001p13348,Sefrioui:2003p13945} but to the best of our knowledge it is presently not fully understood. Fig.~\ref{Mag_SL_VSM}~$(b)$ shows a corresponding M-H magnetization loop of SL-288 at 5\,K that was obtained after the subtraction of the diamagnetic signal of the LSAT substrate. It shows that the saturation magnetization of the LCMO layers reaches a value corresponding to about 2.4\,$\mu_\mathrm{B}$ per Mn ion. Depending on the contribution of the superconducting screening currents, the true average moment per Mn ion may be even slightly higher.\\
      \begin{figure} [b!]
       \begin{picture}(175,175)
          \put(-35,0){\rotatebox{0}{\includegraphics[height=175\unitlength]{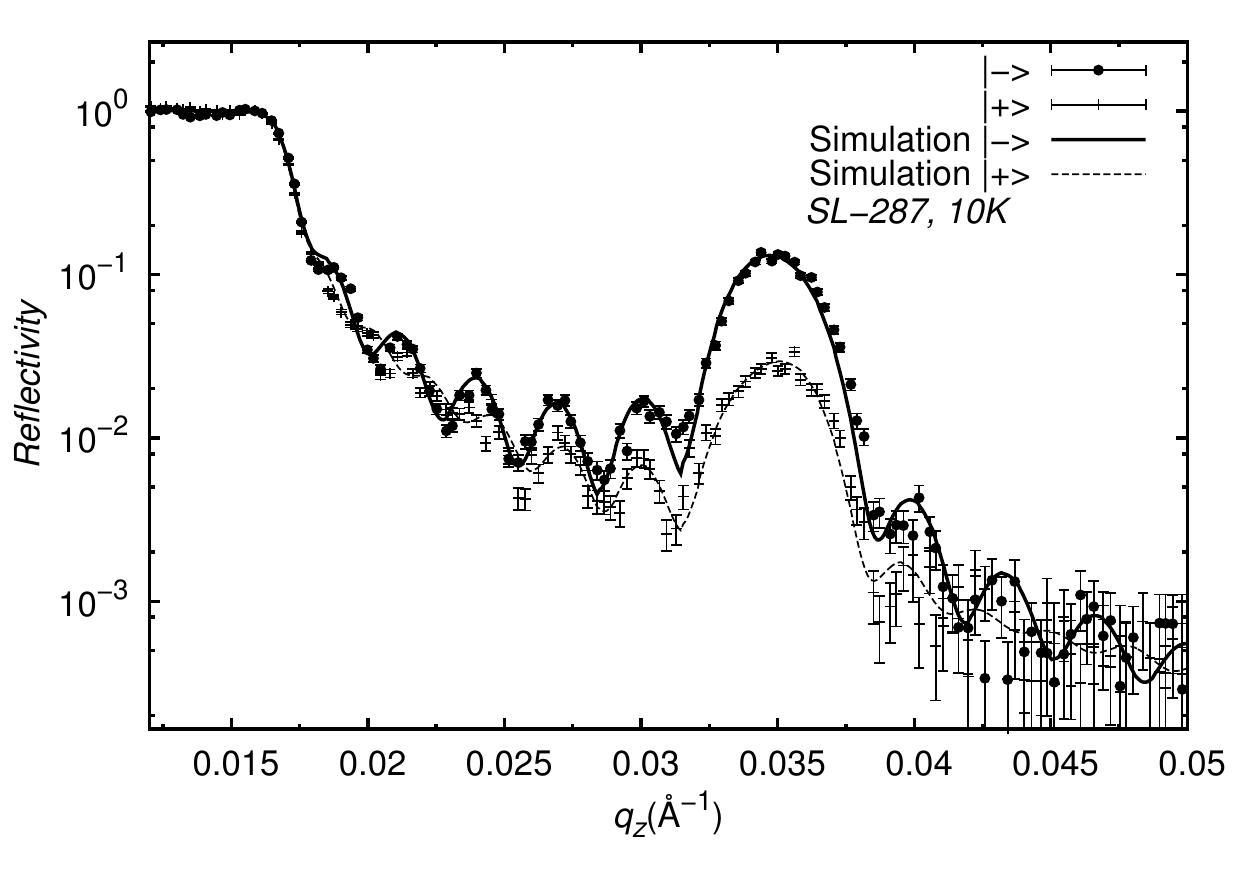}}}
        \end{picture}
        \vspace{0ex}
        \caption{Polarised neutron reflectivity (PNR) curves of SL-287 for spin-up ($|$+$>$) and spin-down ($|$-$>$) polarisation of the neutrons with respect to the direction of the external magnetic field of 1000\,Oe (and thus the mangetisation of the LCMO layers) that is parallel to the layers of the superlattice. The splitting of the first order superlattice peak around $q_\mathrm{z}$=0.35\,\AA$^{-1}$ is a direct measure of the average magnetization density of the superlattice. Assuming that only the Mn ions carry a ferromagnetic moment an average magnitude of 2.7\,$\mu_\mathrm{B}$ per Mn ion has been deduced from the simulation of the spectra with the superfit program as shown by the dashed and solid lines. }
         \label{Mag_SL_PNR} 
      \end{figure}
The magnetic properties of SL-287 which is nominally identical to SL-288 have been investigated with the technique of polarized neutron reflectometry (PNR) as shown in Fig.~\ref{Mag_SL_PNR}. The obtained reflectivity curves for the spin-up and spin-down polarization of the incoming neutrons exhibit a large splitting of the first superlattice Bragg peak at the scattering vector of $q_\mathrm{z}$=0.35\,\AA$^{-1}$ from which the average magnetization density of the LCMO layers can be readily determined. The scattering potential for the polarized neutrons,\,$\rho$\,, is composed of a nuclear contribution,\,$\rho_{nuc}$\,, and a magnetic one,\,$\rho_{mag}$\,, that is proportional to the magnetization density of the sample in the direction parallel to the direction of the neutron spin (which is along the layers of the superlattice). Accordingly, the scattering potential increases for the spin-up polarization of the neutrons while it decreases for the spin-down polarization. Since the nuclear potential of LCMO is smaller than the one of YBCO, the additional magnetic contribution gives rise to a decrease (increase) in the potential difference between the YBCO and LCMO layers. The intensity of the first order superlattice Bragg peak accordingly becomes lower (higher) for the spin up (spin down) neutrons. From the resulting splitting of the spin-up and spin-down PNR curves one can thus deduce the average magnetization density of the superlattice in the direction perpendicular to the scattering vector $q_\mathrm{z}$ (which is parallel to the surface normal of the film) and parallel to the polarization axis of the neutrons (which is along external field and thus parallel to the LCMO layers). With the assumption that only the Mn moments of the LCMO layers carry a ferromagnetic moment, one can thus determine the average magnetic moment per Mn ion. The magnetization density that has been obtained from the simulations of the PNR curves with the superfit program are shown by the solid lines in Fig.~\ref{Mag_SL_PNR}. The deduced magnetic moment amounts to about 2.7\,$\mu_\mathrm{B}$ per Mn ion. This value agrees reasonably well with the 2.4\,$\mu_\mathrm{B}$ per Mn ion that have been obtained from the macroscopic dc magnetization measurements on SL-288. The corresponding magnetization data on SL-287 are less reliable since they are dominated by the large paramagnetic signal of the Nd moments in the NGO substrates. Irrespective of the role of demagnetization effects or superconducting screening effects which may also need to be considered for the interpretation of the magnetization measurements, these results show that the ferromagnetic order in the LCMO layers of these superlattices is fairly strong and almost fully developed. Certainly, the ferromagnetic moments are not strongly reduced as compared to the 3.6\,$\mu_\mathrm{B}$ of bulk LCMO single crystals as it was previously reported from magnetization measurements and PNR measurements for YBCO/LCMO superlattices on SrTiO$_3$ substrates that were grown with a high oxygen pressure sputtering technique\cite{Hoffmann:2005p13352}. Instead our results agree rather well with previous reports on similar PLD grown YBCO/LCMO superlattices on SrTiO$_3$ substrates\cite{Stahn:2005p13350,CHAKHALIAN:2006p3735,Hoppler:2009p18981}.
\section{Conclusions and Summary}
\label{Summ}
In summary, we investigated the structural and the electromagnetic properties of heteroepitaxial [YBa$_2$Cu$_3$O$_7$(n)/La$_{0.67}$Ca$_{0.33}$MnO$_3$(m)]$_\mathrm{x}$ superlattices that were grown with pulsed laser deposition (PLD) on NdGaO$_3$ (110) and Sr$_{0.7}$La$_{0.3}$Al$_{0.65}$Ta$_{0.35}$O$_3$ (LSAT) substrates. The latter are very well lattice matched to YBCO and LCMO with a mismatch of less than 1\% and 0.5\%, respectively. Most importantly, unlike the commonly used SrTiO$_3$ substrates, NGO and LSAT do not exhibit any complex strain effects due to low temperature structural phase transitions whose influence on the electromagnetic properties of the YBCO/LCMO superlattices is difficult to control. The structural properties of our superlattices have been determined with \emph{in-situ} reflection high energy electron diffraction (RHEED) as well as with scanning transmission electron microscopy (STEM), x-ray diffraction and neutron reflectometry. The combined results testify for their high structural quality. They demonstrate that the individual YBCO and LCMO layers are flat and coherent over long lateral distances with a layer thickness that is well controlled on the level of a single monolayer. They also show that the interfaces are sharp and coherent with only few defects due to antiphase boundaries in the YBCO layers that originate from unit cell height steps of the LCMO layers. Our measurements also confirm a previously reported unique layer sequence at the YBCO/LCMO and LCMO/YBCO interfaces where a CuO$_2$ bilayer is situated right at the interface\cite{Varela:2003p17134}. These interfacial CuO$_2$ bilayers are lacking one of their neighboring CuO chain layers and thus half of their charge reservoir. Nevertheless, our resistivity measurements of a n=2 superlattice provide evidence that they remain conducting and even exhibit the onset of a superconducting transition at very low temperature. Accordingly, they suggest that these interfacial CuO$_2$ bilayers do not act as insulating blocking layers that may suppress the proximity coupling between the superconductivity and ferromagnetism in the YBCO and the LCMO layer, respectively.  Our studies also show that a long-term oxygen annealing treatment is required to achieve a full oxygenation of the YBCO layers. This is most likely related to the low density of extended defects, like grain boundaries or screw dislocations, which act as shortcuts for the oxygen diffusion paths and thus effectively reduce the lateral oxygen diffusion length. Finally we performed dc magnetization and polarized neutron reflectometry measurements which establish that the LCMO layers are strongly ferromagnetic with a sizeable average saturation moment of about 2.7$\mu_\mathrm{B}$ per Mn ion.
\section{Acknowledgement}
The work at UniFr has been supported by the Swiss National Science Foundation through grants 200020-11978 and 200020-129484 as well as the NCCR program MaNEP. The PNR experiment has been performed at the Morpheus beamline of the SINQ neutron source of the Paul Scherrer Institut (PSI) in Villigen, Switzerland. The authors thank Masashi Watanabe for the Digital Micrograph PCA plug-in. Work at ORNL (MV) was supported by the Office of Science, Materials Sciences and Engineering Division of the US Department of Energy.  Work at Complutense University was supported by the European Research Council Starting Investigator Award, grant 239739 STEMOX.
\bibliography{VM_PLD}

\end{document}